\documentclass[10pt]{revtex4}
\usepackage{amssymb}
\usepackage{latexsym}
\usepackage{epsfig}
\usepackage{amsmath}
\begin{document}
\title{Bouncing universe of entropy-corrected Friedmann equations}
\author{Amin.Salehi\footnote{
Email:salehi.a@lu.ac.ir} and Mojtaba.Mahmoudi fard}
\address{  Department of physics, Lorestan university,Khoramabad, Iran\\
}

\begin{abstract}
The investigation of quantum gravity effects in order to avoid the big bang singularity is a requisite, so that  the idea of oscillating universes is introduced as an alternative for standard cosmological model. Therefore, the Friedmann equations arising from the quantum corrections of entropy-area relation provide the possibility of studying the oscillating universes. Addition to necessity of surveying the bouncing solutions, however, the values of unknown constant parameters are still a matter of open debate. In this paper, we consider modified Friedmann equations with logarithmic
entropy corrected to evaluate singular-free cosmology solutions for any value of so-
called constant pre-factors and all kinds of   curved universes. The results are argued using the dynamical system techniques and by employing the phase plane analysis for full classification of the nonsingular evolutions.


\end{abstract}

\maketitle

\section{Introduction}

General relativity predicts a spacetime singularity for the beginning of the universe \cite{Riess, Coh}, as it is no longer valid and so there are some alternative theories to standard big bang cosmology \cite{Li, Huang}. Many attempts have been done to resolve this singularity problem through the modified general relativity theory \cite{Hsu}, although it may be resolved via string theory and/or loop quantum cosmology \cite{HDE}, as some candidates for a quantum theory of gravity \cite{Setare1, wang0}. In this right, the scenario of an oscillating universe is to avoid the big bang singularity and replace it with a cyclical evolution \cite{sal}.

The quantum phenomenon of Hawking radiation \cite{Shey} indicates that black hole has a temperature proportional to its surface gravity and an entropy proportional to its horizon area \cite{Shey, Suss1}. This issue led people to consider a profound connection between gravity and thermodynamics which was first addressed by Jacobson \cite{Xin} who disclosed that the Einstein  gravitational theory for the spacetime metric can be derived from the horizon entropy-area realation by using the fundamental Clausius relation  $\delta Q = T\delta S$ \cite{Horava}. The investigations on the relation between Einstein field equations and first law of thermodynamics in the setup of black hole spacetime, have been generalized to the cosmological context to derive Friedmann equations with any spatial curvature by applying the Clausius relation to apparent horizon of the FRW universe \cite{Fischler}-\cite{cai2}. See \cite{cc}  for further studies  of thermodynamical aspects of gravity and \cite{Singh}-\cite{cycl19}  to refer to a recent review on possible cyclic models.

The so-called entropy-area formula which holds only in Einstein gravity has to be modified when some higher order curvature term appears in discussing the quantum correction to black hole
 entropy \cite{cai2},\cite{Gao}-\cite{Chevallier}. Therefore, the singularity problem come from a ``shortage of time'' in the early universe should be resolved and replaced by a quantum bounce \cite{Jassal}.

Inclusion of quantum effects, motivated from the loop quantum gravity (LQG) due to thermal
equilibrium and quantum fluctuations \cite{Wetterich, Ma}, leads to the curvature correction in the Einstein-Hilbert action \cite{Granda}   provided as logarithmic entropy-corrected \cite{Daly}-\cite{Kom2}

\begin{eqnarray}\label{action}
S=\frac{A}{4G}+\alpha\ln\frac{A}{4G}+\beta\frac{4G}{A}
\end{eqnarray}

in entropy-area relationship of black holes in classical gravity, where $A=4\pi \tilde{r}_{A}^{2}$ which $\tilde{r}_{A}=\frac{1}{\sqrt{H^{2}+\frac{k}{a^{2}}}}$ is the radius  of apparent horizon,  $\alpha$ and $\beta$ are dimensionless constants. The exact values of these constants are not yet determined and is still in debate within the quantum geometry  (LQG). Various approaches to the black hole entropy yield the logarithmic correction involving  $-\frac{1}{2}$  or $-\frac{3}{2}$ as popular values of  $\alpha$  coefficient \cite{Kaul}. In these treatments, there is no such consensus with regard to how one might fix the value of the logarithmic pre-factor $\alpha$ (e.g. with $\beta=0$), as it appears to be a highly model dependent parameter \cite{Gour}.

Accordingly, for the time being the aim of this paper is twofold. The first is to see the importance of quantum effects on modification of Friedmann equations as to investigate the possibility of having solutions of the singularity-free cosmological model.  The other is to evaluate the values of the constant pre-factors of equation (1) at bouncing evolution, corresponding to occurrence of bounce and having oscillating solutions, since they are not yet determined even within the loop quantum gravity. For instance, some works lead to negative or positive  $\alpha$  \cite{Kaul}-\cite{37}; and some reference argued that  $\alpha$  should be equal to zero \cite{37}.

The paper is organized as follows. In the next section we briefly discuss the modified Friedmann equations  and bouncing approach by the action of the corrected entropy-area relation (1) to the apparent horizon of FRW universe. In section III we get a spacetime of cosmological perturbations by studying the structure of the dynamical system via Jacobian stability combined with phase plane analysis. Some realizations of cosmological bounces and oscillating behaviors are discussed in section IV. This situation, in general,  is argued  for all values of debated parameters $\alpha$ and $\beta$ by employing for different curvatures, as shown in Table (1). We conclude in Sec. V with a discussion of approach and some key results for  bouncing  evolution of the entropy-corrected Friedmann equations on the metric of FRW universe.

\section{Modified Friedmann equations}

Addition to thermodynamical law of black hole's entropy as a dominant term given by a quarter of its horizon area, there is a quantum correction involving the logarithmic of the area \cite{mit}. On this basis, by tacking the corrected entropy-area relation (1) into account, the modified Friedmann equations of a Friedmann-Robertson-Walker (FRW) universe  with any spatial curvature whose matter content
constitutes a cold dark matter with the density $\rho$, would be \cite{cai2}

\begin{eqnarray}\label{fried1}
H^{2}+\frac{k}{a^{2}}+\frac{\alpha G}{2\pi}(H^{2}+\frac{k}{a^{2}})^{2}-\frac{\beta G^{2}}{3\pi^{2}}(H^{2}+\frac{k}{a^{2}})^{3}=\frac{8\pi G}{3}\rho,
\end{eqnarray}
\begin{eqnarray}\label{fried2}
2(\dot{H}-\frac{k}{a^{2}})(1+\frac{\alpha G}{\pi}(H^{2}+\frac{k}{a^{2}})-\frac{\beta G^{2}}{\pi^{2}}(H^{2}+\frac{k}{a^{2}})^{2})=-8\pi G (\rho+p),
\end{eqnarray}
where we put  $G=1$; and $k=0, 1, -1$ represent a flat, closed and open universe, respectively. We also get  $\frac{\alpha G}{\pi}$  as a new $\alpha$ and  $\frac{\beta G^{2}}{\pi^{2}}$  as a new  $\beta$ in the next sections of paper for simplicity. Note that, given the recent astrophysical data \cite{w18}, the universe is undergoing a state of accelerating expansion. We will thus suppose that dark energy might play an important role on the implications for the fate of the universe, and it's matter content  is dominated by a dust matter (with the pressure $p=0$) at early time \cite{cycl19}.

Later on in the following sections within the dynamical calculation of these equations we shall see that the stability of nonsingular solutions occurs according to results summarized in table (1) and depicted in phase plane portraits corresponding to particular regions/values of parameters $\alpha$ and $\beta$ for different curvature of matter dominated universes.

\section{Stability analysis}

In order to investigate bouncing evolution of universe, the stability structure of dynamical systems can be studied combined with phase plane analysis, by introducing the following new variables:

\begin{eqnarray}
\chi=H\  ,  \ \zeta=a   ,\ \ \eta=\rho
\end{eqnarray}
From equations (\ref{fried1}) and (\ref{fried2}), the evolution equations of these variables become,
\begin{eqnarray}\label{xdot}
\dot{\chi}&=&-\frac{(1+\gamma)\eta}{2(1+2\alpha^{'}(\chi^{2}+k\zeta^{-2})-3\beta(\chi^{2}+k\zeta^{-2})^{2})}+\frac{k}{\zeta^{2}}\\
\dot{\zeta}&=&\zeta\chi\label{ydot}
\end{eqnarray}
Where a dot denotes a derivative with respect to the cosmic
time. From equation (\ref{fried1}), $\eta$ will be obtained as
\begin{eqnarray}\label{eta}
 \eta=3(\chi^{2}+k\zeta^{-2}+\alpha(\chi^{2}+k\zeta^{-2})^{2}-\beta(\chi^{2}+k\zeta^{-2})^{3})
\end{eqnarray}
For next a bounce can also be defined locally. The minimal conditions from a local point of view for a bounce to happen in the case of a FLRW universe were analyzed in
\cite{Paris}, where a Tolman wormhole was defined as a universe that undergoes a collapse, attains
a minimum radius, and subsequently expands. Thus, to have a bounce it is necessary that $\dot{a}_{b}=0 $ and $\ddot{a}_{b} >0$, or equivalently $\chi_{b}=0$ and $\frac{d\chi}{dt}|_{t_{b}}>0$ are satisfied. For simplicity, we define $h_{b}=\frac{d\chi}{dt}|_{t_{b}}$; as from equation (\ref{xdot})
\begin{equation}\label{xb}
h_{b}=-\frac{k(\zeta_{b}^{4}-\alpha k\zeta_{b}^{2}+3k^{2}\beta)}{2\zeta_{b}^{2}(\zeta_{b}^{4}+2\alpha k\zeta_{b}^{2}-3k^{2}\beta)}
\end{equation}

and from equation (\ref{eta}), one can obtain the energy density at bounce, as

\begin{equation}\label{xcd}
\rho_{b}=\frac{3k(\zeta_{b}^{4}+\alpha\zeta_{b}^{2}-\beta)}{\zeta_{b}^{6}}
\end{equation}

It is also of interest to have a bounce without violation of Null Energy Condition, so that it requires
\begin{equation}\label{cond}
 h_{b}>0 \ \ \ , \ \ \ \rho_{b}>0)
\end{equation}
Note that although the satisfaction of conditions (\ref{cond}) is necessary for a cyclic universe scenario
in which the universe oscillates through a series of expansions and contractions, but it also needs more conditions from the dynamical system techniques.

Here, we  employ the dynamical system technique to investigate when the system can oscillate, displaying that how a system behaves near a critical point. Under which condition the system can oscillate near the critical points. The phase plane analysis addresses the stability of solutions and  trajectories of dynamical systems under small perturbations of initial conditions. Therefore,  it can generally
give us a full picture of the dynamics combined with phase plane analysis, especially in cosmology. Since such a study can be used to circumvent the need for initial conditions which is as one of the shortcomings of the standard cosmological model (SCM) \cite{sal, cycl5}. In this right, the phase plane analysis is an invaluable tool in studying and visualizing the behavior of dynamical systems. The phase portraits in configuration space display the certain characteristics of system if the dynamics are stable or not. Hence one can know how trajectories behave near the critical points, e.g. whether they move toward or away from the fixed points \cite{43-45}.

As a well-studied problem, therefore, the stability of an orbit in phase space is reduced to a certain characteristics of system associated with the eigenvalues of Jacobian.
In this case there are  four critical points in the phase space,  extreme points to describe the
asymptotic behavior of the system, that are determined by simultaneously solving $\chi'=0$  and $\zeta'=0$.

\begin{equation}
\chi_{1c}=0,\ \ \zeta_{1c}=\sqrt{\frac{k}{2(1+3\gamma)}}\Big(-3\gamma\alpha-\alpha+(9\gamma^2\alpha^2-6\gamma \alpha^2+\alpha^2-24\gamma\beta-12\beta+36\gamma^2\beta)^{\frac{1}{2}}\Big)^{\frac{1}{2}}
\end{equation}
\begin{equation}
  \chi_{2c}=0,\ \ \zeta_{2c} =-\sqrt{\frac{k}{2(1+3\gamma)}}\Big(-3\gamma\alpha-\alpha+(9\gamma^2\alpha^2-6\gamma \alpha^2+\alpha^2-24\gamma\beta-12\beta+36\gamma^2\beta)^{\frac{1}{2}}\Big)^{\frac{1}{2}}
\end{equation}
\begin{equation}
\chi_{3c}=0,\ \ \zeta_{3c}=\sqrt{\frac{-k}{2(1+3\gamma)}}\Big(-3\gamma\alpha-\alpha+(9\gamma^2\alpha^2-6\gamma \alpha^2+\alpha^2-24\gamma\beta-12\beta+36\gamma^2\beta)^{\frac{1}{2}}\Big)^{\frac{1}{2}}
\end{equation}
\begin{equation}
\chi_{4c}=0,\ \ \zeta_{4c}=-\sqrt{\frac{-k}{2(1+3\gamma)}}\Big(-3\gamma\alpha-\alpha+(9\gamma^2\alpha^2-6\gamma \alpha^2+\alpha^2-24\gamma\beta-12\beta+36\gamma^2\beta)^{\frac{1}{2}}\Big)^{\frac{1}{2}}
\end{equation}

By virtue of this framework, we can study the stability analysis of critical points to
investigate the properties of the dynamical system (Eqs. (5) and (6)).  Considering that a nonlinear autonomous system may be approximated by a linear system through its coefficient matrix (Jacobian), one can obtain the eigenvalues of Jacobian to standard classification of the different types of the fixed points \cite{sal}. In fact, the critical points of a system can be almost completely classified based on their eigenvalues, e.g. if they are complex, the point is called a focus and if system have eigenvalues with real part zero, the critical point is called center which is a neutrally stable closed orbit.

Here, the stability analysis is carried out for matter dominated model $\gamma=0$. Evaluating the Jacobian at the critical points gives us the eigenvalues of system to each of them as follows,

\begin{equation}\label{lambda}
\lambda_{\pm}=\pm \frac{k\sqrt{3}}{3\zeta_{c}(\alpha\zeta_{c}^{2}-2\beta k)}\Omega^{\frac{1}{2}}
\end{equation}\\
Where
\begin{equation}\label{omeg}
\Omega=-\alpha^{3}\zeta_{c}^2+3k\alpha^{2}\beta+8\alpha \zeta_{c}^2\beta-12k\beta^{2}
\end{equation}\\
The parameter $\Omega$ is an important characteristic which its sign determines the nature of  critical points and consequently explains the behavior of  system near them. We shall obtain more insight into its dynamics in the next section.

\section{Oscillating evolution of curved universes}

In the following, we focus on matter dominated models to represent the oscillating evolution of different curved universes. The phase plane diagram of pertinent systems and in some cases the cyclic behavior of the scale factor are illustrated in the field plots of phase space, that gives us the opportunity to study all of the evolution paths admissible for all initial conditions. Therefore, the bouncing trajectories are shown in some detail and different oscillating solutions  are obtained by tacking a certain condition for each portrait in the configuration space of this bouncing cosmology scenario.

\subsection{Case of spatially flat ($k=0$) universe}

In the case of the cold dark matter dominated model, the system has an infinite number of critical points on $\chi=0$-axis.  Also, there is  four critical points on  $\zeta=0$ line in phase space as follows:
\begin{equation}
\zeta_{1c}=0,\ \ \chi_{1c}=\frac{\sqrt{2}}{2\beta}\sqrt{\beta(\alpha+\sqrt{\alpha^{2}+4\beta}}
\end{equation}
\begin{equation}
 \zeta_{2c}=0,\ \ \chi_{2c}=-\frac{\sqrt{2}}{2\beta}\sqrt{\beta(\alpha+\sqrt{\alpha^{2}+4\beta}}
\end{equation}
\begin{equation}
\zeta_{3c}=0,\ \ \chi_{3c}=\frac{\sqrt{2}}{2\beta}\sqrt{\beta(\alpha-\sqrt{\alpha^{2}+4\beta}}
\end{equation}
\begin{equation}
\zeta_{4c}=0,\ \ \chi_{4c}=-\frac{\sqrt{2}}{2\beta}\sqrt{\beta(\alpha-\sqrt{\alpha^{2}+4\beta}}
\end{equation}

   All of the critical points on $\chi=0$ line, have the same eigenvalue as $\lambda_{\pm}=(0,0)$. However, each of the critical points on $\zeta=0$ line, have the eigenvalue $\lambda_{\pm}=(-3\chi_{c},\chi_{c})$ in the phase space. Since the eigenvalues are real, the universe does not oscillate. Considering that the minimal conditions require
a bounce with ($ H_{b}=0,\dot{H}_{b}>0$) that are evaluated at the bounce, the equation (\ref{fried2}) implies that the bounce condition can not be satisfied. In fact, even if bounce occurs the null energy condition will be violated $(\rho+p)<0$ \cite{53-55}, so that in this case there is no bouncing solution. \\

   \subsection{Case of hyperspherical ($k=1$) universe}

In this case there are in general four critical points in the phase space
as follows:
\begin{equation}
P_{1}=\chi_{c}=0,\zeta_{c}=\frac{\sqrt{2}}{2}\sqrt{\alpha+\sqrt{\alpha^{2}-12\beta}},\ \ \
P_{2}=\chi_{c}=0,\zeta_{c}=-\frac{\sqrt{2}}{2}\sqrt{\alpha+\sqrt{\alpha^{2}-12\beta}},\\ \nonumber
\end{equation}
\begin{equation}
P_{3}=\chi_{c}=0,\zeta_{c}=\frac{\sqrt{2}}{2}\sqrt{\alpha-\sqrt{\alpha^{2}-12\beta}},\ \ \
P_{4}=\chi_{c}=0,\zeta_{c}=-\frac{\sqrt{2}}{2}\sqrt{\alpha-\sqrt{\alpha^{2}-12\beta}},\\ \nonumber
\end{equation}

 where they are mirror images of each other along the $\zeta=0$ line, considering that the condition  $\alpha^{2}>12\beta$ must be satisfied. The number of critical points and their properties are based on their eigenvalues, that they depend on the value of $\alpha$ and $\beta$ parameters. In the following, the respective classification of critical points in the phase space and corresponding eigenvalues are obtained for oscillating solutions accompanied by phase plane trajectories of system. Also, from equations (\ref{xb}) and (\ref{xcd}) $h_{b}$ and $\rho_{b}$ would be
 \begin{equation}\label{xb2}
h_{b}=-\frac{(\zeta_{b}^{4}-\alpha \zeta_{b}^{2}+3\beta)}{2\zeta_{b}^{2}(\zeta_{b}^{4}+2\alpha \zeta_{b}^{2}-3\beta)}
\  \ , \ \ \rho_{b}=\frac{3(\zeta_{b}^{4}+\alpha\zeta_{b}^{2}-\beta)}{\zeta_{b}^{6}}
\end{equation}

\subsubsection{$k=1,\alpha=positive,\beta=positive$}

In this case $\alpha>\sqrt{\alpha^2-12\beta}$ is preserved, so that through the condition  $\alpha^{2}>12\beta$, all of the four critical points exist.The eigenvalues of $P_{3}$ and $P_{4}$ are real with positive sign thus their natures are saddle, however the corresponding eigenvalues of $P_{1}$ and $P_{2}$ are purely imaginary. To do so, the relation $\zeta^{2}_{c}=\frac{1}{2}(\alpha+\sqrt{\alpha^{2}-12\beta})$ holds for $P_{1}$ and $P_{2}$ points, thus by substituting it in the equation (\ref{omeg}),the parameter $\Omega$ will be

\begin{equation}
\Omega=\Big(-\frac{1}{2}\alpha^{4}-\frac{1}{2}\alpha^{3}\sqrt{\alpha^2-12\beta}+4\alpha\beta\sqrt{\alpha^2-12\beta}+7\alpha^{2}\beta-12\beta^{2}\Big)
\end{equation}

which is simplified as

\begin{equation}\label{lambda2}
\Omega=\Big(-\frac{1}{2}\{(\alpha^2-12\beta)(\alpha^2-2\beta)+(\alpha^2-8\beta)\sqrt{\alpha^2-12\beta}\}\Big)
\end{equation}

 For positive values of  $\alpha$ and $\beta$ and via the condition $(\alpha^2-12\beta>0)$, automatically the terms of $(\alpha^2-2\beta)$ and $(\alpha^2-8\beta)$ are positive. Therefore, $\Omega$ is negative and the eigenvalues are imaginary with zero real parts. This implies that the nature of the critical points $P_{1}$ and $P_{2}$ is center and marginally stable. Hence, curves in the phase space are closed trajectories around the center. Interestingly, this behavior in phase space, indicates that the scale factor of the universe undergoes contracting and expanding
phases periodically, so that the universe can possess an
exactly cyclic evolution.
 Also from equation (\ref{xcd}), it directly follows that  for positive values of $\alpha$ and $\beta$ ,the energy density  would be positive at bounce only if,
 \begin{equation}\label{condition1}
\zeta_{b}>\frac{1}{2}\sqrt{-2\alpha+2\sqrt{\alpha^{2}+4\beta}}
\end{equation}
 Thus, from all bouncing oscillating  trajectories of the phase space, those trajectories  which satisfy the condition (\ref{condition1}) are admissible solutions. To clarify this, we have plotted the phase space $a-H$ for $\alpha=1$ and $\beta=0.05$ (Fig. (1)). Also evaluation of scale factor and energy density as a function of time for important trajectories  have been shown in Fig. (2). By assuming  $\alpha=1$ and $\beta=0.05$, the four critical points would be

\begin{equation}
P_{1}=(\chi=0,\zeta=0.9),\ \ P_{2}=(\chi=0,\zeta=-0.9) ,\ \ P_{3}=(\chi=0,\zeta=0.4),\ \ P_{4}:(\chi=0,\zeta=-0.4)\nonumber
\end{equation}

Evaluating the Jacobian at the critical points, we get

\begin{equation}
\lambda_{1}=(0.5I,\zeta=-0.5I),\ \ \lambda_{2}=(0.5I,\zeta=-0.5I),\ \ \lambda_{3}=(1.6,-1.6),\ \ \lambda_{4}=(1.6,-1.6) \nonumber
\end{equation}

Where they are corresponding eigenvalues, so that the stable critical points $P_{1}$ and $P_{2}$ are called center, while $P_{3}$ and $P_{4}$ are  unstable fixed points or  saddle points singularity. It would be interesting to see whether the initial conditions are important in cosmological dynamics systems. From equation (\ref{condition1}), for $\alpha=1$ and $\beta=0.05$, if $a_{b}=\zeta_{b}>0.23$, one can see then the energy density will be positive.  Fig. (2) also illustrates  the  point that the two curves (red and blue), where both have approximately the same bounce but a little difference in initial conditions,  have made completely different evolution in their energy density. The red curve, with positive energy density  which  $a_{b}=\zeta_{b}>0.23$,   satisfies the condition (\ref{condition1}), while the blue curve $a_{b}=\zeta_{b}<0.23$   does not.

\begin{tabular*}{2.5 cm}{cc}
\includegraphics[scale=.42]{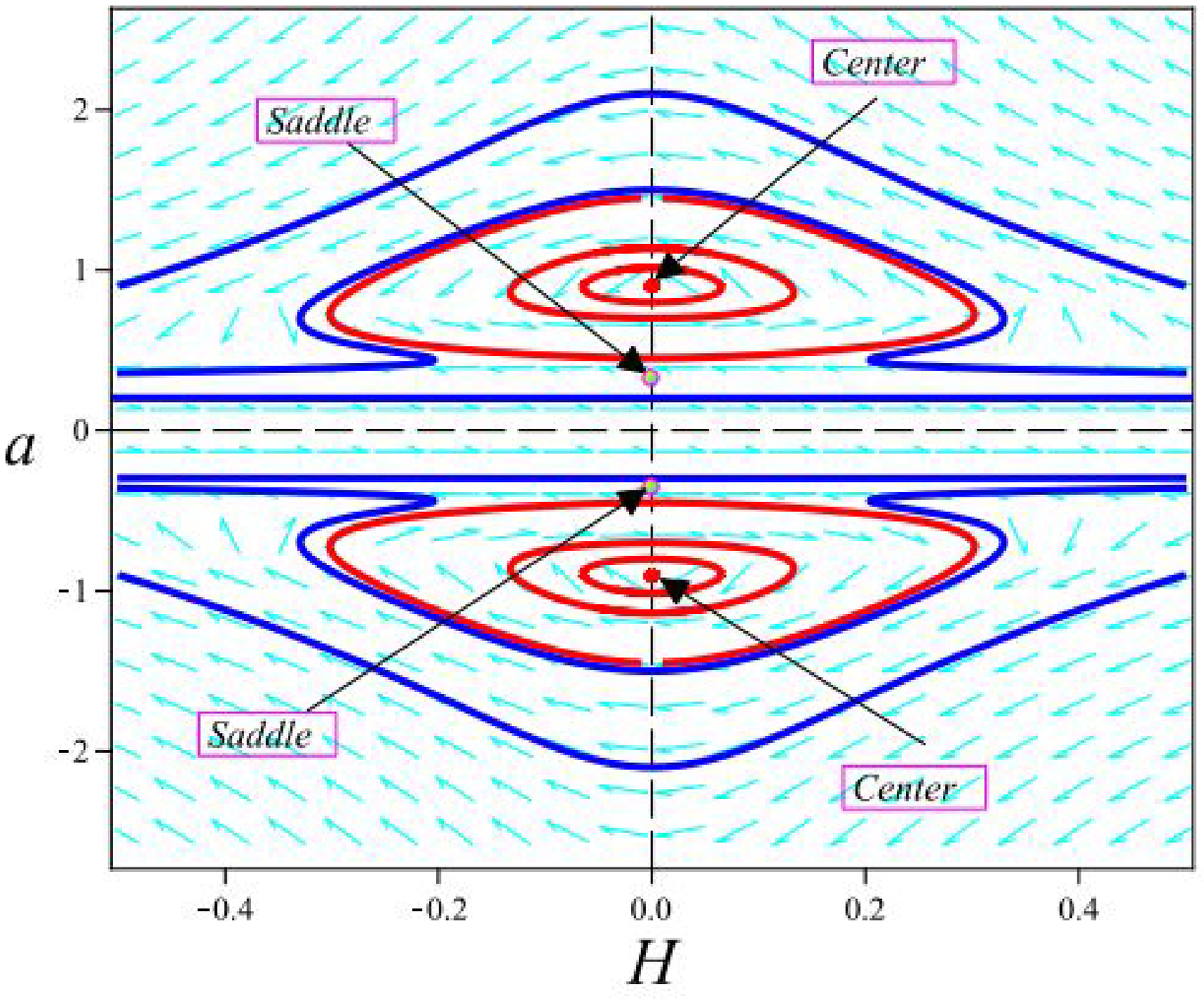}\hspace{0.1 cm} \includegraphics[scale=.31]{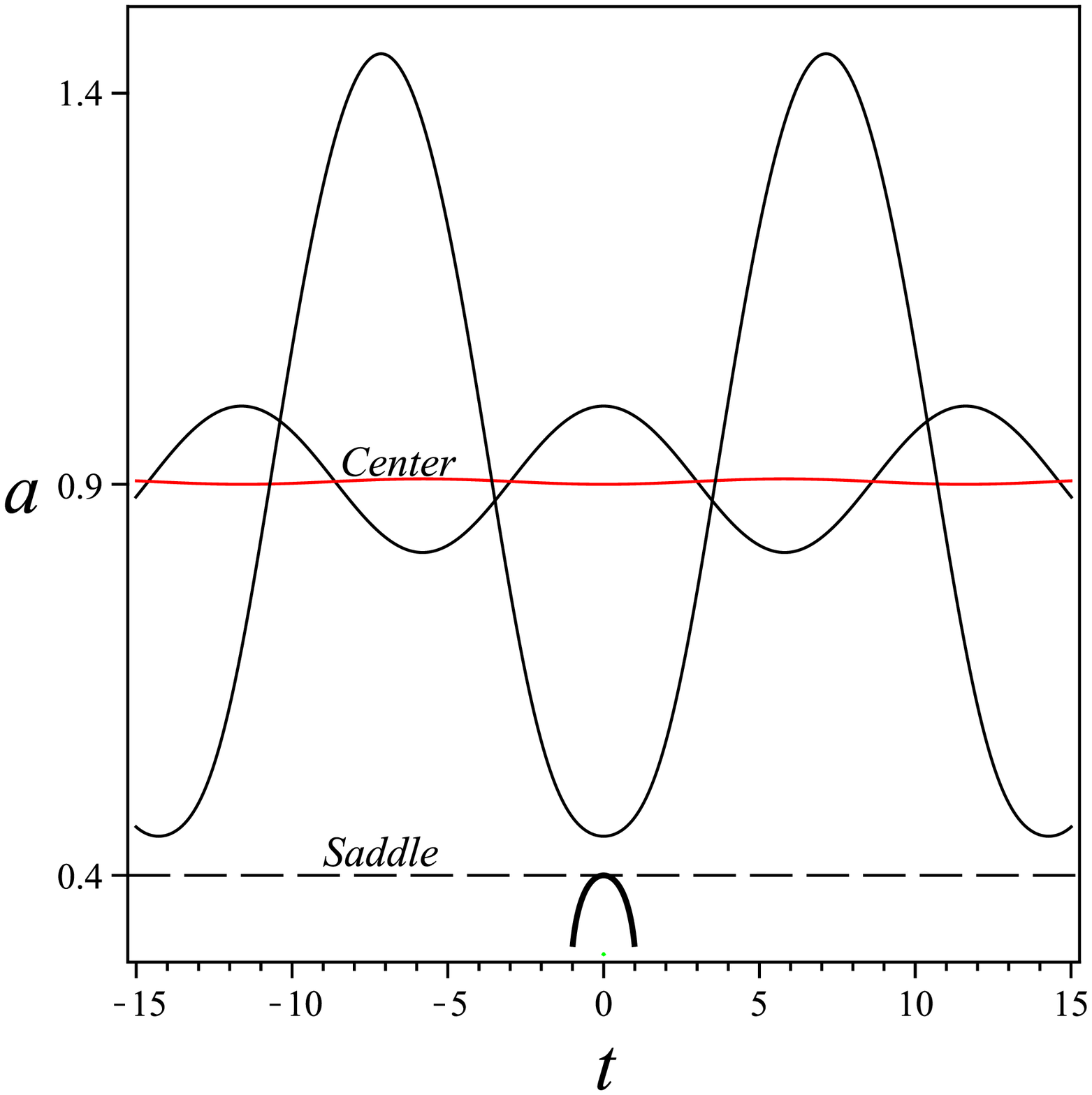}\hspace{0.1 cm} \\
Fig.1. (Left) Dynamical behavior of the system around the critical points.\\ (Right) Time evolution of the scale factor $a$ corresponding to the trajectories in phase space \\for the case of $\gamma=0$, $k=1$, $\alpha=1$ and $\beta=0.05$.
\end{tabular*}\\

\begin{tabular*}{2.5 cm}{cc}
\includegraphics[scale=.3]{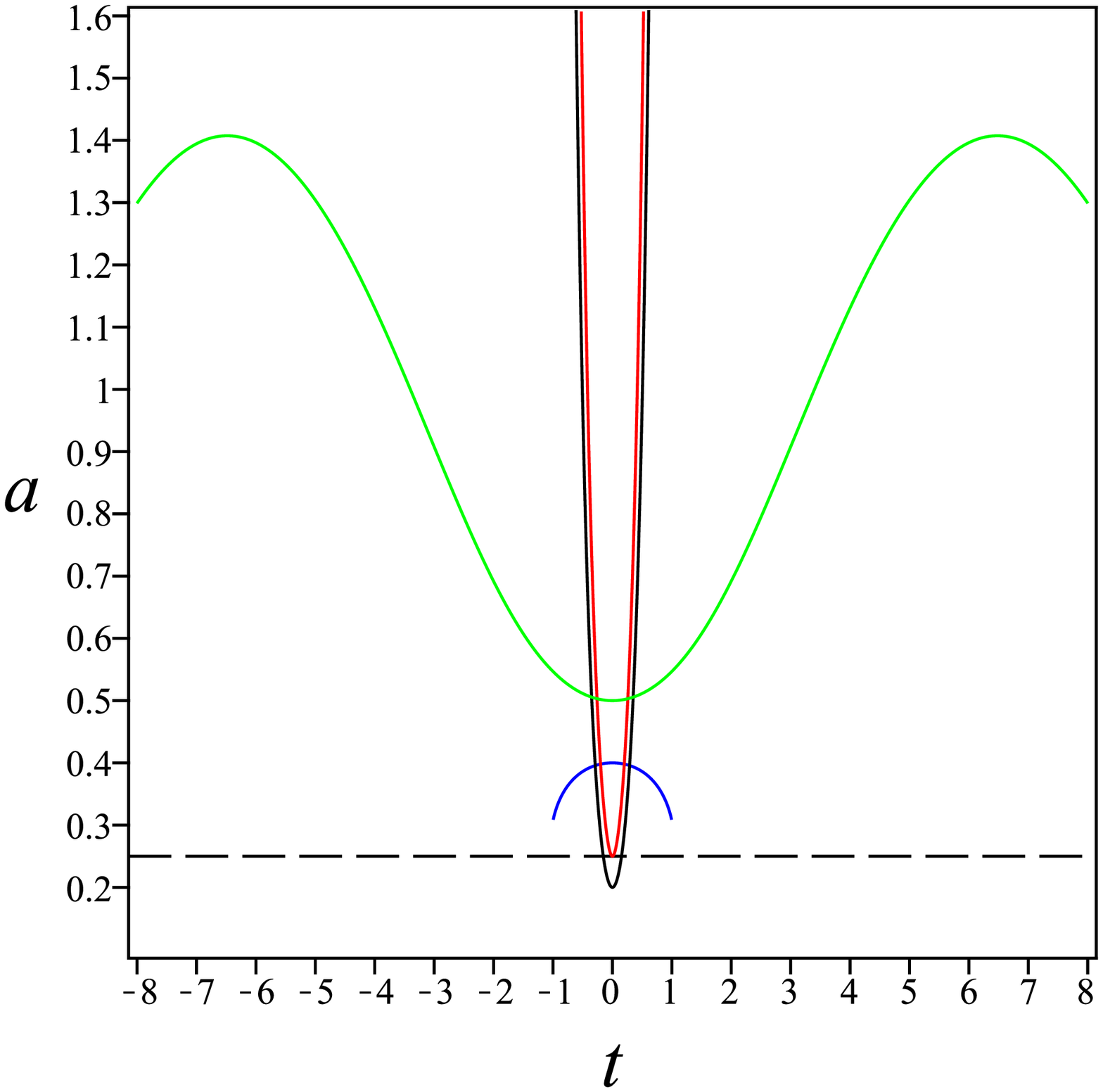}\hspace{0.1 cm} \includegraphics[scale=.3]{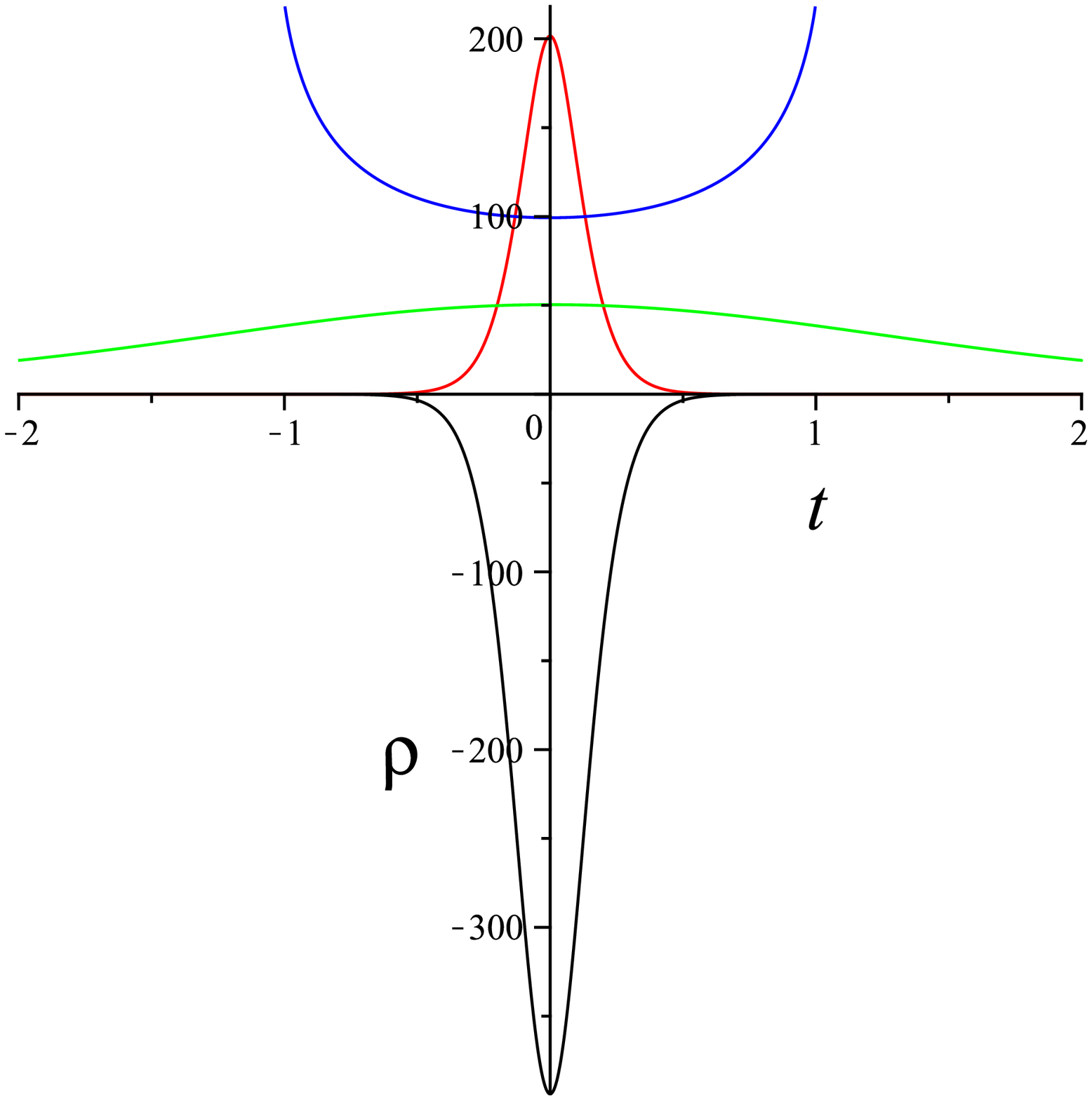}\hspace{0.1 cm} \\
Fig.2 Time evolution of the scale factor $a$ and energy density $\rho$, corresponding to the  different initial conditions in\\ phase space   for the case of $\gamma=0$, $k=1$, $\alpha=1$ and $\beta=0.05$.
\end{tabular*}\\

\subsubsection{$k=1,\alpha=positive,\beta=negative$}

For positive $\alpha$ and negative $\beta$, $\alpha<\sqrt{\alpha^2-12\beta}$ is preserved, and the terms of $(\alpha^{2}-12\beta)$, $(\alpha^{2}-2\beta)$ and $(\alpha^{2}-8\beta)$ are positive.
Therefore, there are only two of the above critical points, $P_{1}$ and $P_{2}$, which again are centers. The  corresponding eigenvalues are also purely imaginary and the universe has an oscillating behavior in the phase space.
 As shown in Fig. (3), for example, by setting $\alpha=1$ and $\beta=-1$, the two critical points would be $P_{1}=(\chi=0,\zeta=1.5)$ and $P_{2}=(\chi=0,\zeta=-1.5) $ with the same eigenvalue
$\lambda_{1,2}=(0.5I,-0.5I)$

\begin{tabular*}{2.5 cm}{cc}
\includegraphics[scale=.42]{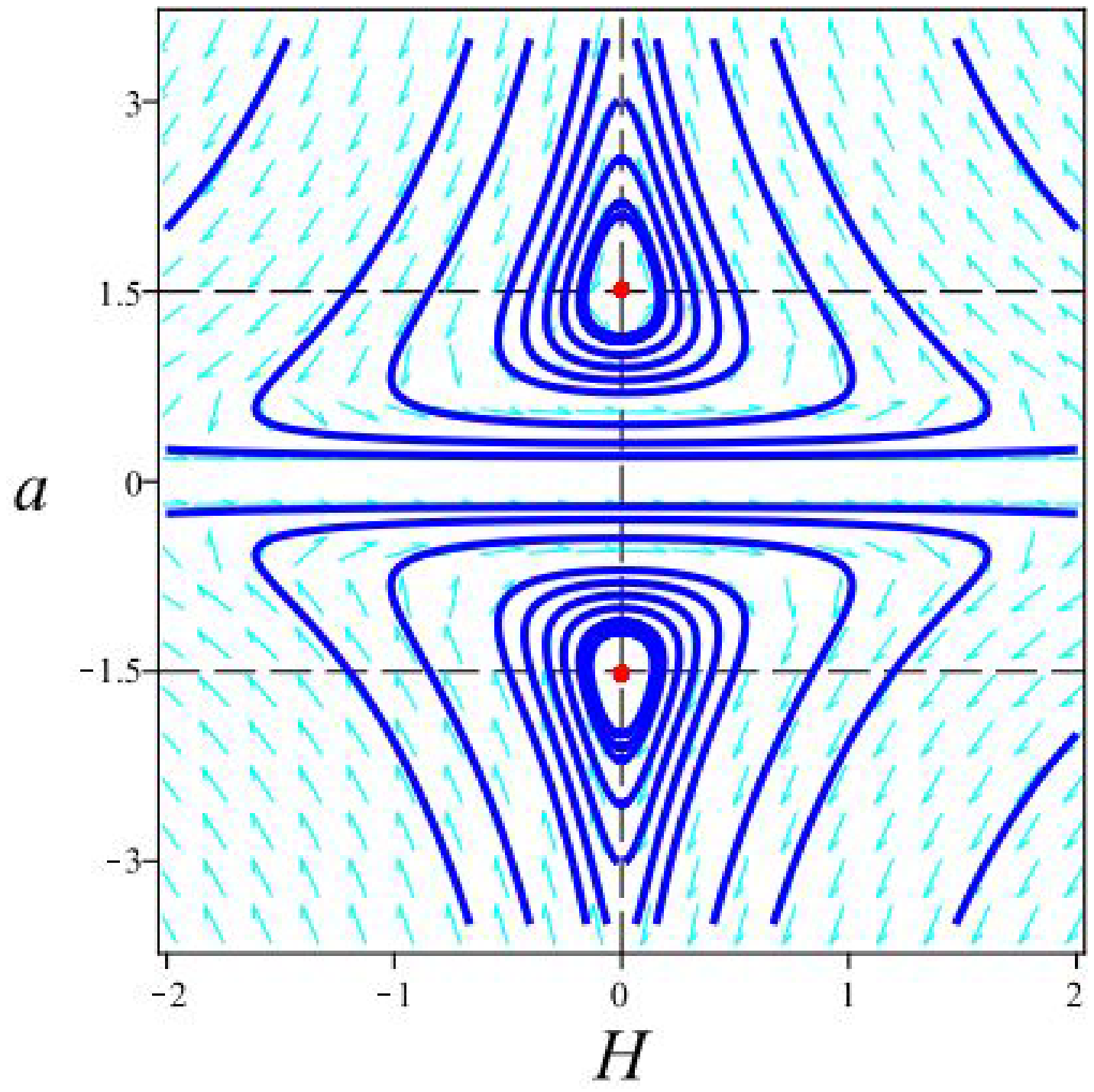}\hspace{0.1 cm} \includegraphics[scale=.32]{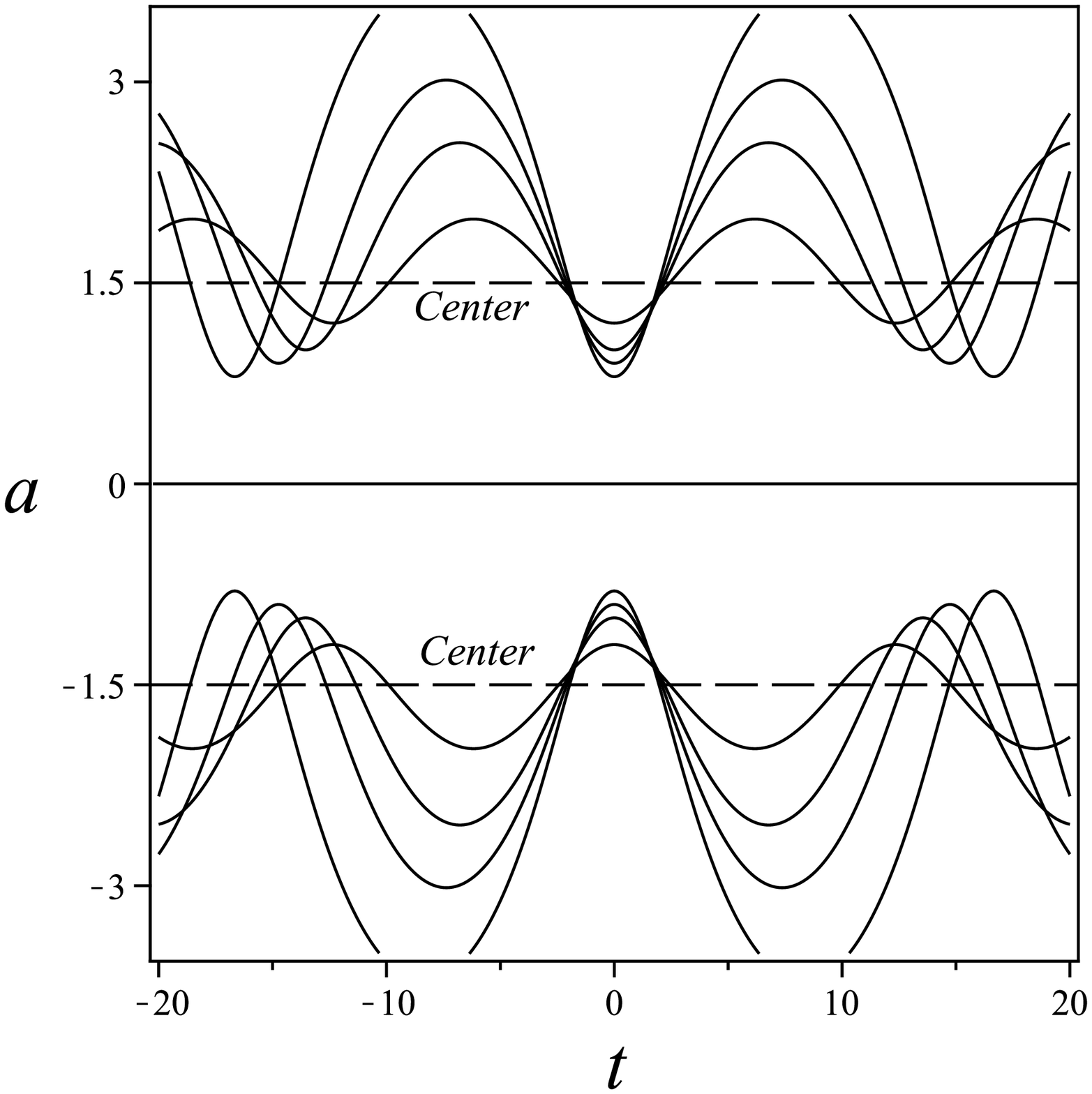}\hspace{0.1 cm} \\

Fig.3. (Left) Dynamical behavior of the system around the critical points. \\(Right) Time evolution of the scale factor $a$ corresponding to the trajectories in phase space \\for the case of $\gamma=0$, $k=1$, $\alpha=1$ and $\beta=-1$.
\end{tabular*}\\

One may note that from equation (\ref{condition1}), the energy density at bounce point in this case is positive.

\subsubsection{$k=1,\alpha=negative,\beta=positive$}

 Since $\alpha$ is negative and $\beta$ is positive, the expressions $\alpha-\sqrt{\alpha^2-12\beta}<0$ and $\alpha+\sqrt{\alpha^2-12\beta}<0$ are established; so that the system does not have any real critical point and consequently the oscillating solutions do not exist. However, a single
bounce without oscillation can occur under the appropriate
conditions. This is possible to make by satisfaction of minimal requirement of bounce condition $(H_{b}=0 $ and $\dot{H}_{b} >0)$, and it can also follow equivalently at the bounce, as $\frac{d\chi}{dt}|_{t_{b}}>0$ in terms of new variables.

  In order to explain this more fully, it is useful to obtain the energy density and $(\frac{d\chi}{dt})$ at bounce. From equation (\ref{eta}),  the energy density will be

   \begin{equation}\label{xcdd}
\rho_{b}=\frac{3(\zeta_{b}^{4}+\alpha\zeta_{b}^{2}-\beta)}{\zeta_{b}^{6}}
\end{equation}

   This relation indicates that only under the condition $\Big(\zeta_{b}>\frac{\sqrt{2}}{2}\sqrt{-\alpha+\sqrt{\alpha^{2}+4\beta}}\Big)$, the energy density will be positive. Also, from equation (\ref{xdot})

   \begin{equation}\label{xb}
h_{b}=\frac{d\chi}{dt}|_{t_{b}}=-\frac{(\zeta_{b}^{4}-\alpha\zeta_{b}^{2}+3\beta)}{2\zeta_{b}^{2}(\zeta_{b}^{4}+2\alpha\zeta_{b}^{2}-3\beta)}
\end{equation}

Since the positive energy density automatically implies the positive $(\zeta_{b}^{4}+\alpha\zeta_{b}^{2}-\beta)$, the following inequality holds for this case

\begin{equation}
\zeta_{b}^{4}-\alpha\zeta_{b}^{2}+3\beta>(\zeta_{b}^{4}+\alpha\zeta_{b}^{2}-\beta)
\end{equation}

Therefore,  the positive energy density also naturally implies the positive $\zeta_{b}^{4}-\alpha\zeta_{b}^{2}+3\beta$, as from equation (\ref{xb}), the bounce would be occur only for negative  $\zeta_{b}^{4}+2\alpha\zeta_{b}^{2}-3\beta$ or equivalently when $\Big(\zeta_{b}<\sqrt{-\alpha+\sqrt{\alpha^{2}+3\beta}}\Big)$. One can   conclude  that under  following condition

\begin{equation}\label{in2}
(\frac{\sqrt{2}}{2}\sqrt{-\alpha+\sqrt{\alpha^{2}+4\beta}}\Big)<\zeta_{b}<(\sqrt{-\alpha+\sqrt{\alpha^{2}+3\beta}}\Big)
\end{equation}

the bounce occurs and energy density will be positive. It is interesting to note that, it is possible to having a bounce while the energy density is negative  only when the r.h.s of above inequality (\ref{in2}) is just preserved.
 The phase plane diagrams for the case of $\alpha=-1$, $\beta=1$ have been plotted in Fig. (4) for clarifying this point, as the evolution of scale factor  and energy density for different initial conditions $\zeta_{b}$ have also been shown.

In accordance to equation (\ref{in2}), $\frac{\sqrt{2}}{2}\sqrt{1+\sqrt{5}}\simeq 1.27<\zeta_{b}<\sqrt{3}\simeq1.7$ to avoid a negative energy density at bounce, eventually,   there exists  bouncing solution with positive energy density.\\
\begin{tabular*}{2.5 cm}{cc}
\includegraphics[scale=.3]{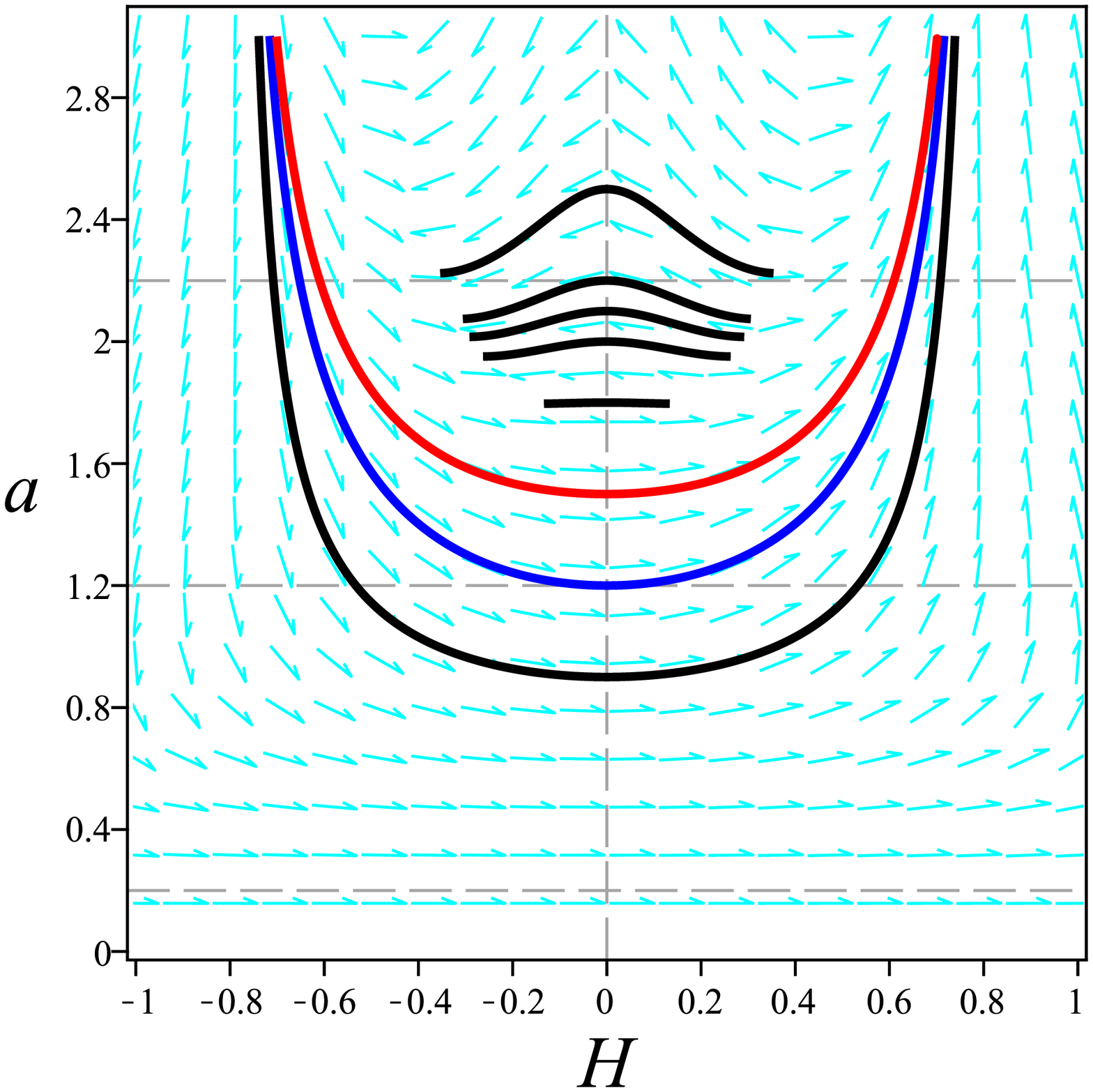}\hspace{0.1 cm}
\includegraphics[scale=.3]{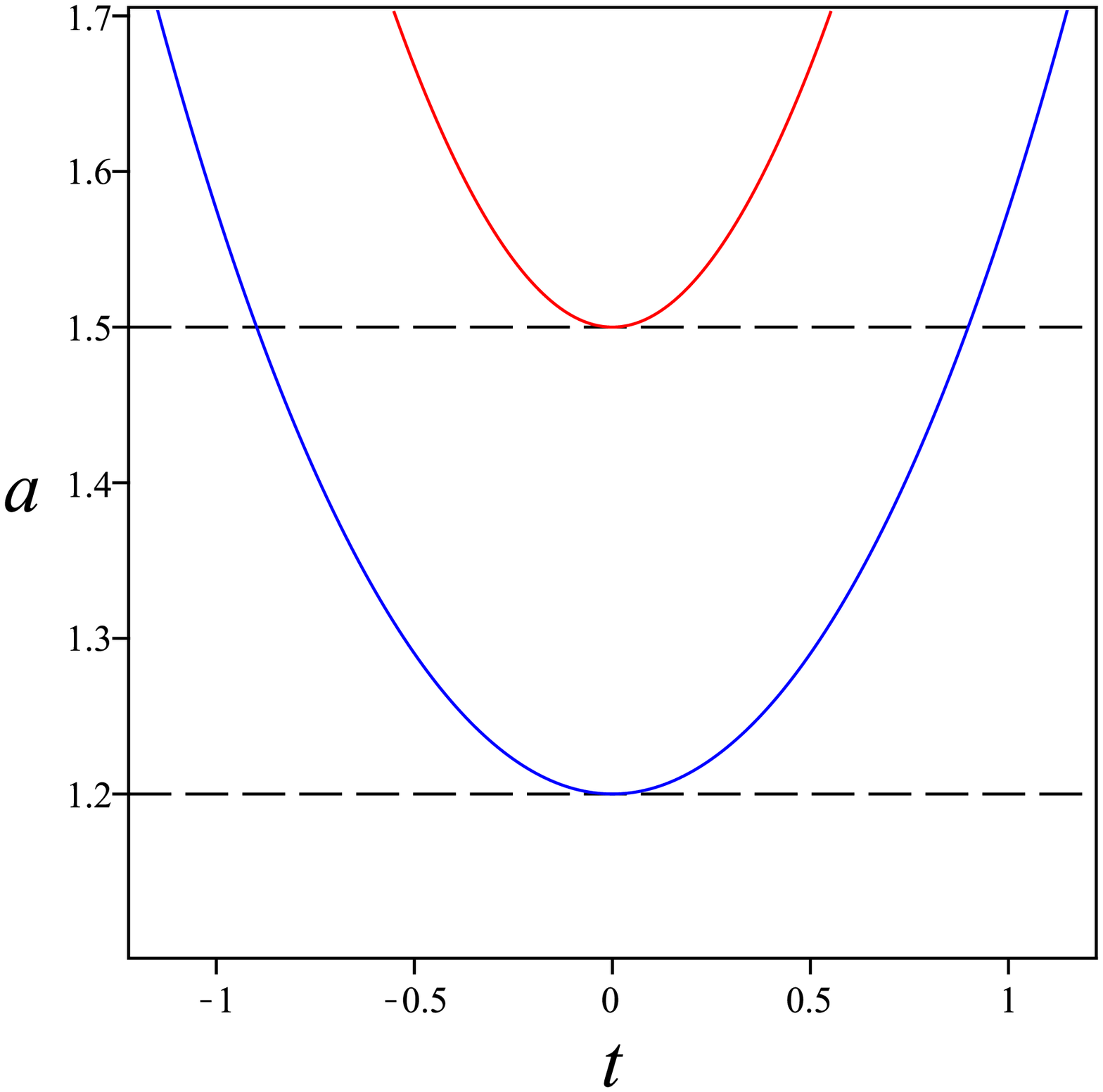}\hspace{0.1 cm}
 \includegraphics[scale=.3]{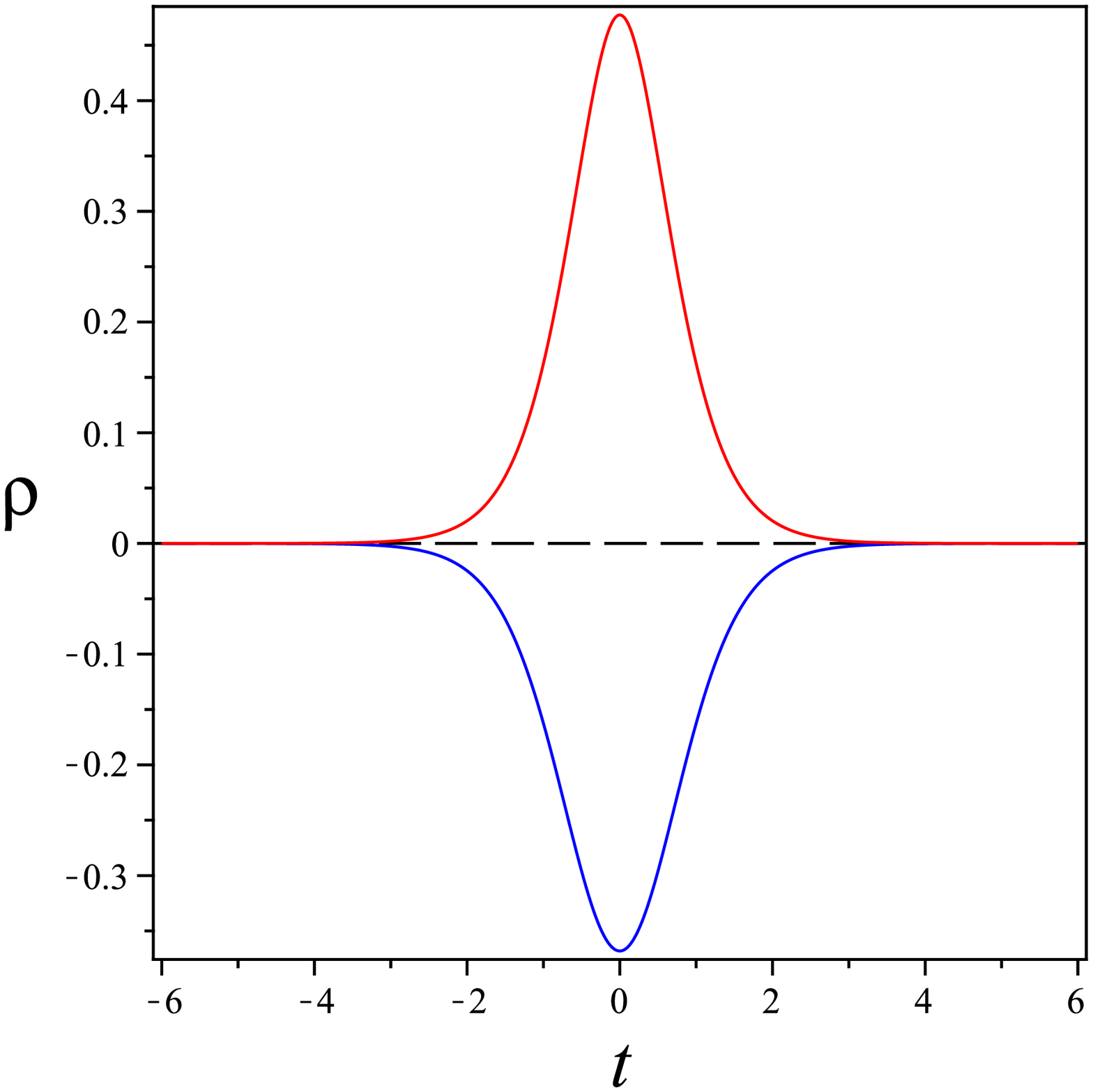}\hspace{0.1 cm}\\
Fig.4.(Left) Dynamical behavior of the system around the critical points\\ (Middle)  Time evolution of the scale factor corresponding to the red and blue trajectories.  \\ (Right) Time evolution of the energy density corresponding to the trajectories of
  the Fig.
\\
\end{tabular*}\\

\subsubsection{$k=1,\alpha=negative,\beta=negative$}

For this case as in the case of  $\alpha>0$, $\beta<0$,   there are only two critical points, $P_{1}$ and $P_{2}$. The three dimensional plot of $\Omega$  as a function of $\alpha$ and $\beta$ has been drawn in Fig. (5), which    in some regions    is positive and (in other parts) negative. Thus the eigenvalues of the system can be real or complex, as in  those regions that $\Omega$ is negative
  the universe has  oscillating solutions  and trajectories are closed  orbits  in the phase space.\\
  \begin{tabular*}{2.5 cm}{cc}
\includegraphics[scale=.42]{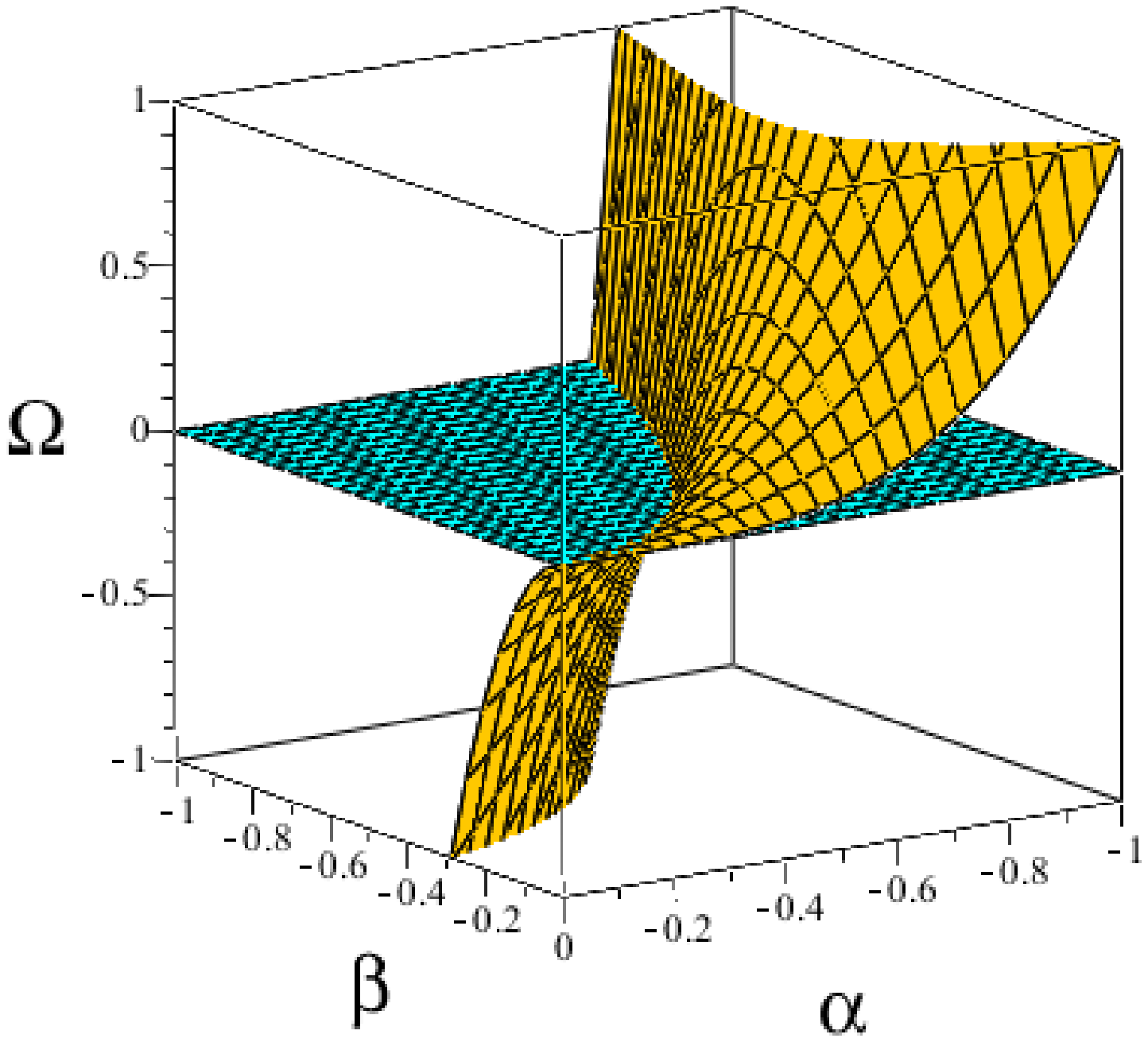}\hspace{0.1 cm}\\
Fig. 5:   The three dimensional   plot of $\Omega$ as a function of $\alpha$ and $\beta$ for $k=1$ \\
\end{tabular*}\\
For instance, let us get $\alpha=-0.1,\beta=-0.1$ according to Fig. (6). Hence, the value of $\Omega$ is negative and the eigenvalues are purely imaginary, as the trajectories in phase plane  are closed
curves  near the critical points.  These points have the characteristics of a
center fixed point and  system oscillates around them. Similar to what was found for the case of positive $\alpha$ and  positive  $\beta$, the energy density would be positive if the condition $\Big(\zeta_{b}>\frac{\sqrt{2}}{2}\sqrt{-\alpha+\sqrt{\alpha^{2}+4\beta}}\Big)$ is satisfied.\\
\begin{tabular*}{2.5 cm}{cc}
\includegraphics[scale=.42]{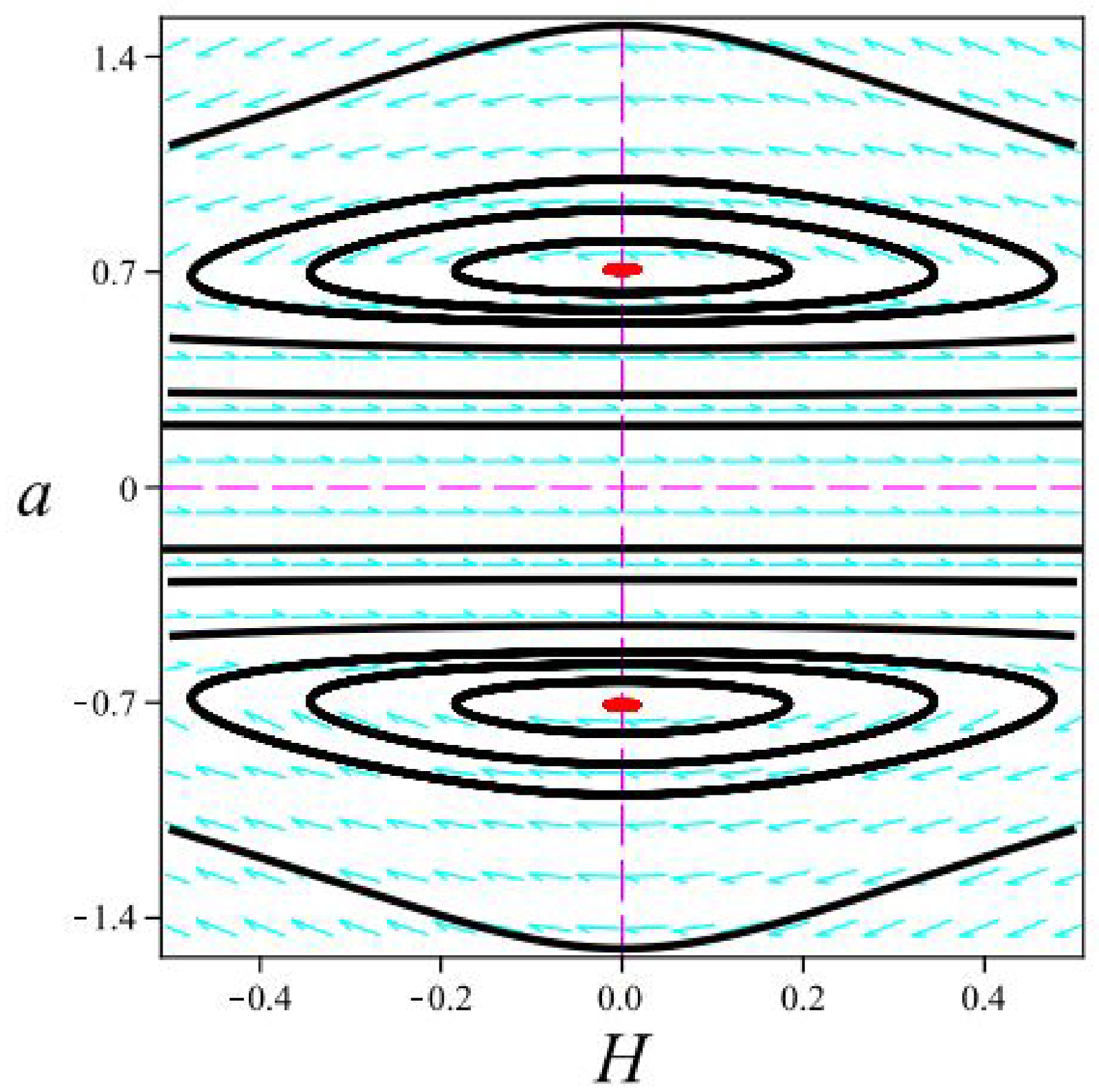}\hspace{0.1 cm}\\
Fig. 6:  Phase plane of  parameters $(a,H)$ for the case of $k=1,\alpha=-0.1,\beta=-0.1$  . \\
\end{tabular*}\\

\subsubsection{$k=1,\alpha=0,\beta=0$}

The system has one critical point $(\chi_{c}=0,\zeta_{c}=0)$ and its eigenvalues are $\lambda_{\pm}=0$,  so that the universe does not oscillate.  Applying equation (\ref{fried1}) into the (\ref{fried2}) yields

\begin{equation}\label{in2s}
2\dot{H}+3H^{2}=-\frac{k}{a^{2}}
\end{equation}

 and thus $\dot{H}$ is negative, as  the universe does not oscillate and bounce never occur. In this case the universe expands and reaches its maximum size and then collapses. The energy density decreases during  the expansion and it reaches its minimum values then increases. The two dimensional phase space of $(a,H)$ and time evolution of scale factor have been shown in Fig. (7).\\\\

\begin{tabular*}{2.5 cm}{cc}
\includegraphics[scale=.42]{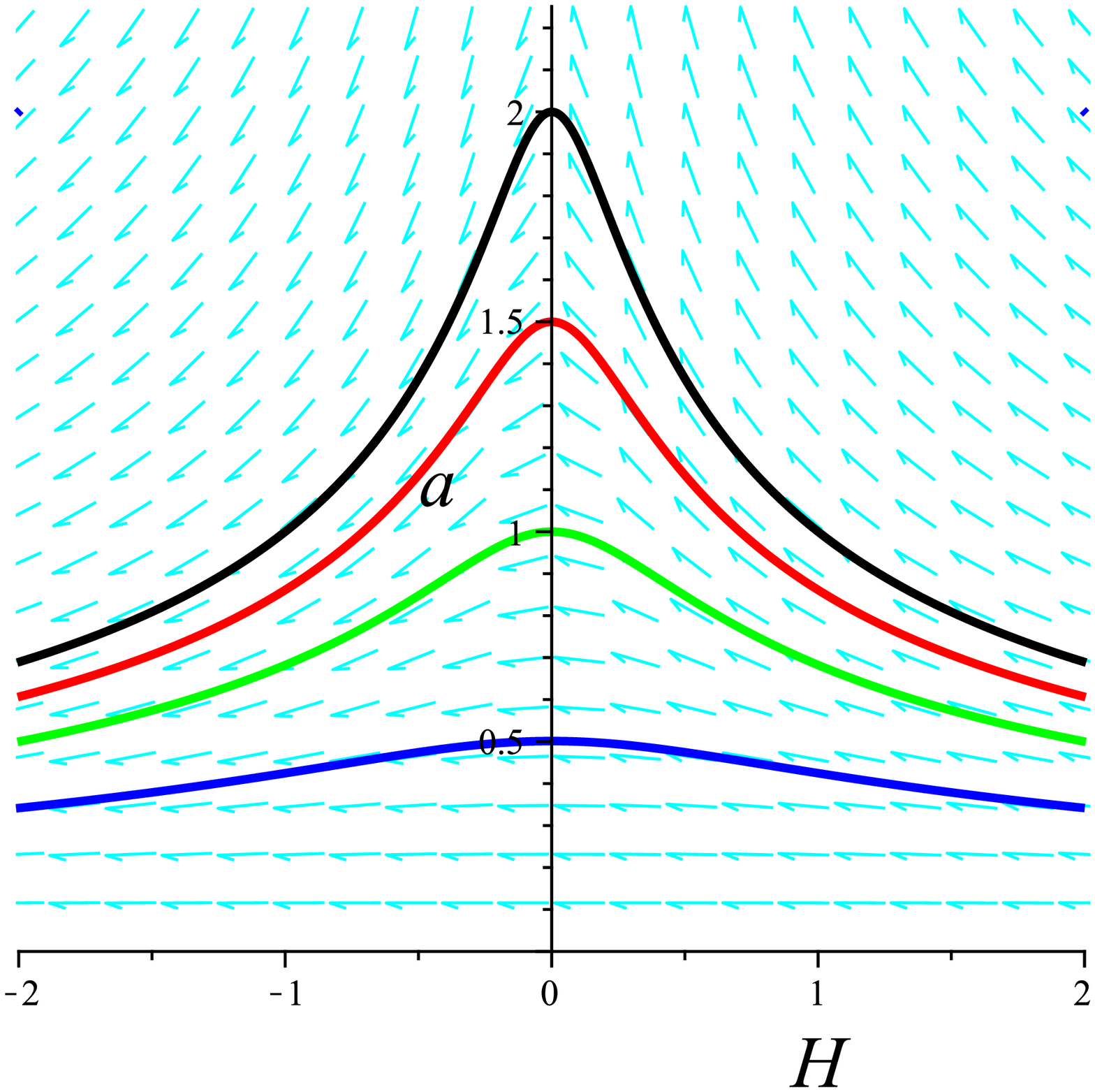}\hspace{0.1 cm}
\includegraphics[scale=.42]{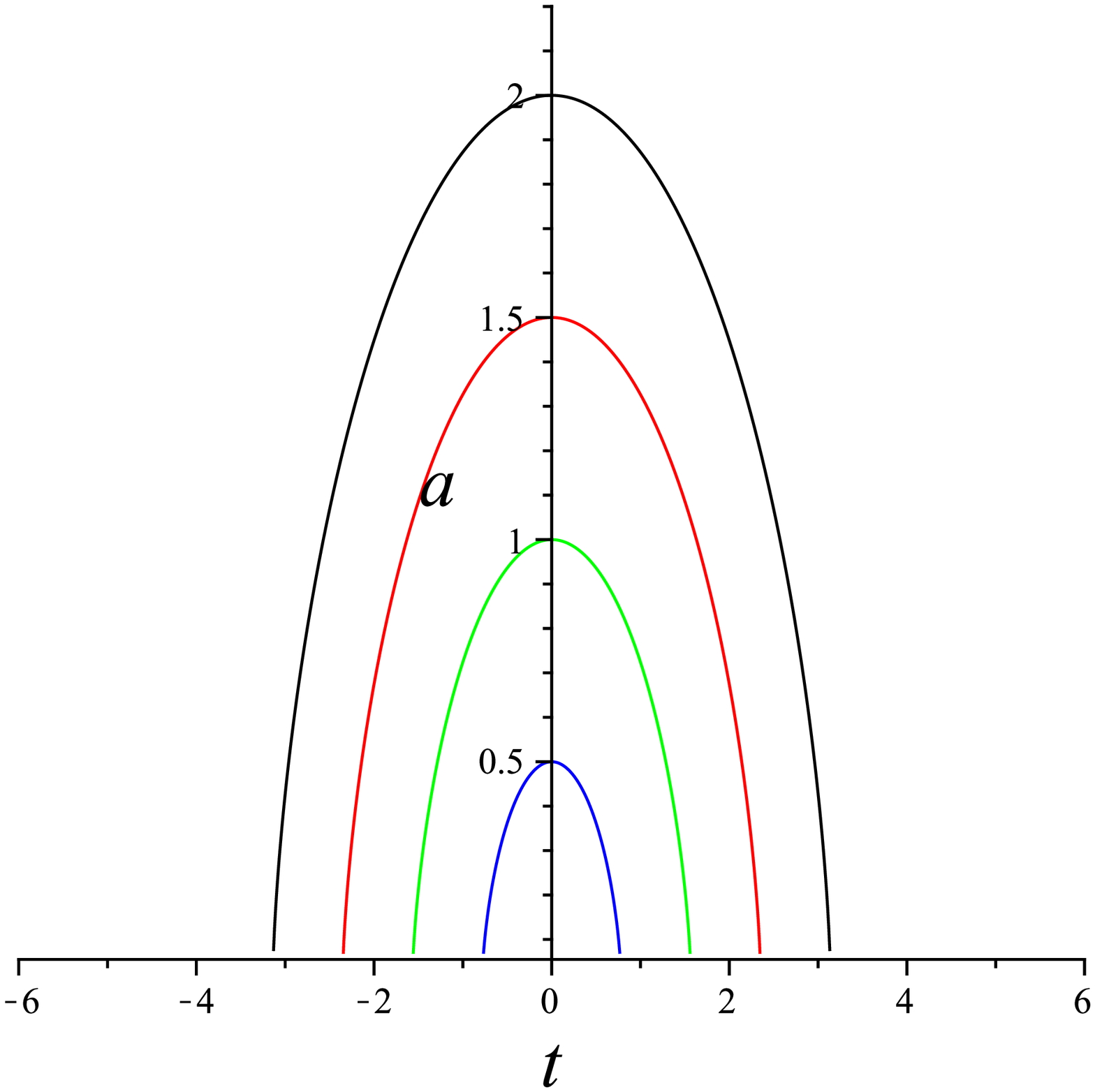}\hspace{0.1 cm}\\
Fig. 7: Left): Phase plane of  parameters $(a,H)$ for the case of $k=1,\alpha=0,\beta=0$\\ Right):Time evolution of the scale factor corresponding to the highlighted trajectories of
  the phase space.
 \\
\end{tabular*}\\

\subsubsection{$k=1,\alpha=positive,\beta=0$}

For this case the system has two critical points $(\chi_{c}=0,\zeta_{c}=\sqrt{\alpha})$ and $(\chi_{c}=0,\zeta_{c}=-\sqrt{\alpha})$ , the   corresponding eigenvalues are $\lambda_{1}=(\frac{i\sqrt{3}}{3\sqrt{\alpha}},\frac{-i\sqrt{3}}{3\sqrt{\alpha}})$ and $\lambda_{2}=(\frac{i\sqrt{3}}{3\sqrt{\alpha}},\frac{-i\sqrt{3}}{3\alpha})$  which are the same.
The eigenvalues are purely imaginary and conjugated, as the critical points are centers (nonhyperbolic critical points) with closed curves turns around them.
This evolution predicts a cyclic universe where the minimal and maximal
values of the scale factor remain the same in every
cycle (see Fig. (8)).  \\\\

\begin{tabular*}{2.5 cm}{cc}
\includegraphics[scale=.42]{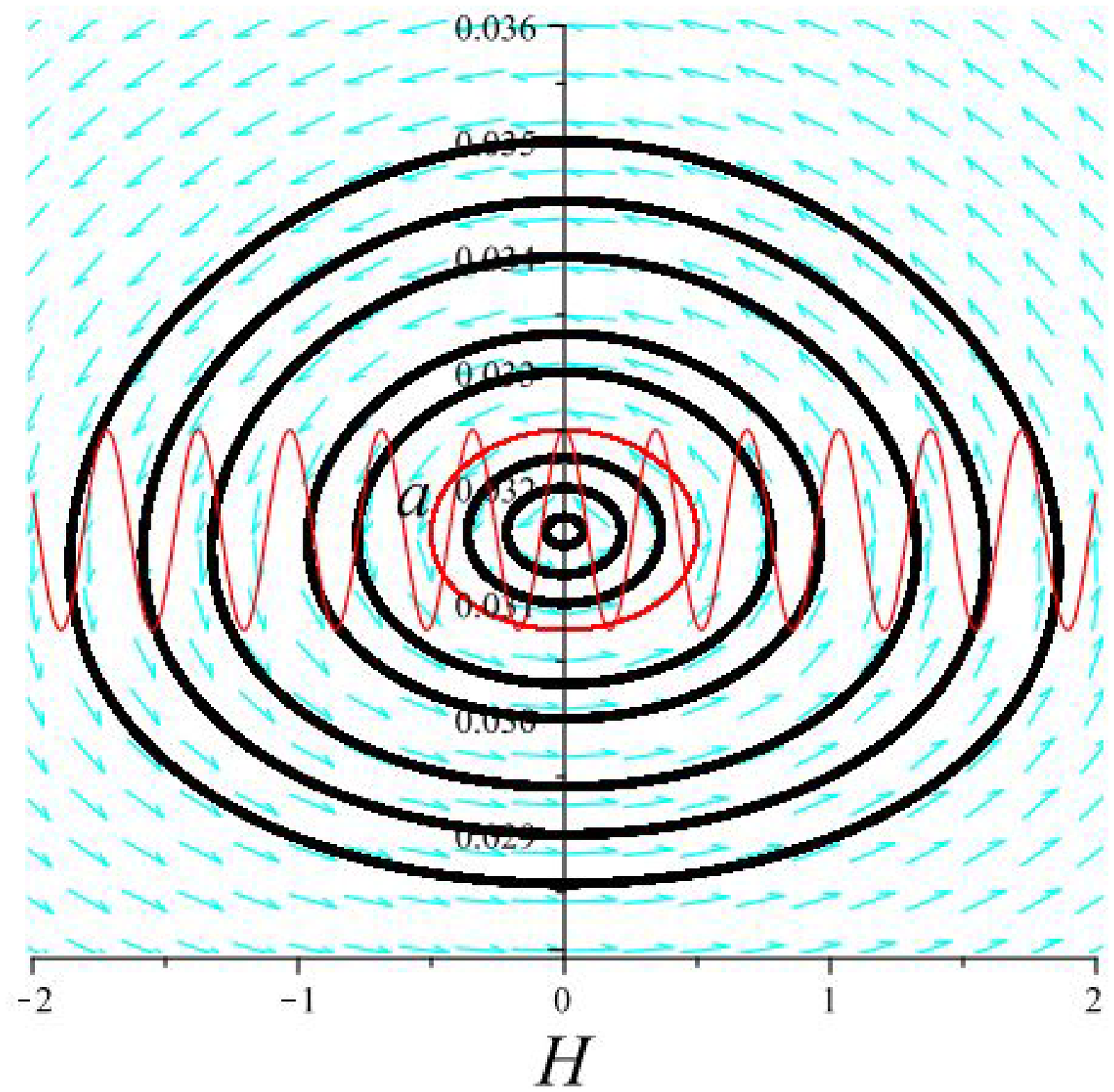}\hspace{0.1 cm}\\
Fig. 8:Phase space of the dynamical system for the case  $k=1,\alpha=0.001,\beta=0$. The dynamical behavior of the\\ scale factor for an arbitrary  trajectory of phase space (red trajectory) has been plotted \\
\end{tabular*}\\
Moreover, the equation (\ref{xcd}) implies that   for positive values of $\alpha$ and $\beta=0$, the energy density at bounce is positive.

\subsubsection{$k=1,\alpha=negative,\beta=0$}

In this case, the system doesn't have any real critical point and it doesn't oscillate, however  minimal  condition for a bounce $\chi_{b}=0$ and $\frac{d\chi}{dt}|_{t_{b}} >0$ may be satisfied. If we put $\chi=0$ in r.h.s of equation (\ref{xdot}), the expression  $\frac{d\chi}{dt}=-\frac{\zeta^{2}-\alpha}{\zeta^{2}(\zeta^{2}+2\alpha)}$ is obtained. This condition denotes that for $\zeta_{b}=a_{b}<\sqrt{-2\alpha}$, the bounce can occur and from equation (\ref{xcd}) the energy density will be positive. In order to illustrate  this matter in phase space configuration, by setting  $\alpha=-2$ as shown in Fig. (9). For $a_{b}<2$ the bounce take places whereas for $a_{b}>2$, the bounce backs or collapses. \\
\begin{tabular*}{2.5 cm}{cc}
\includegraphics[scale=.3]{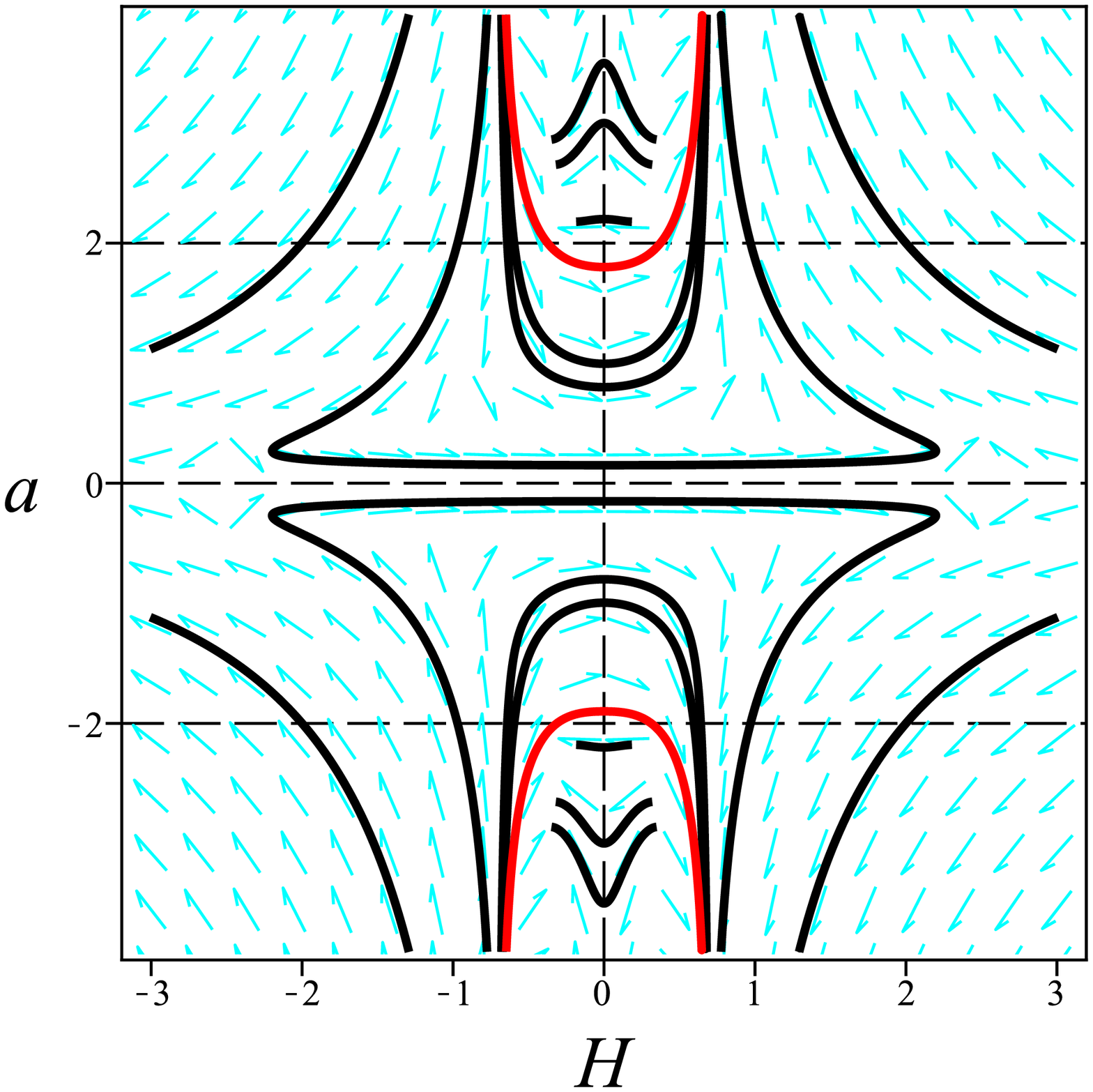}\hspace{0.1 cm}\includegraphics[scale=.3]{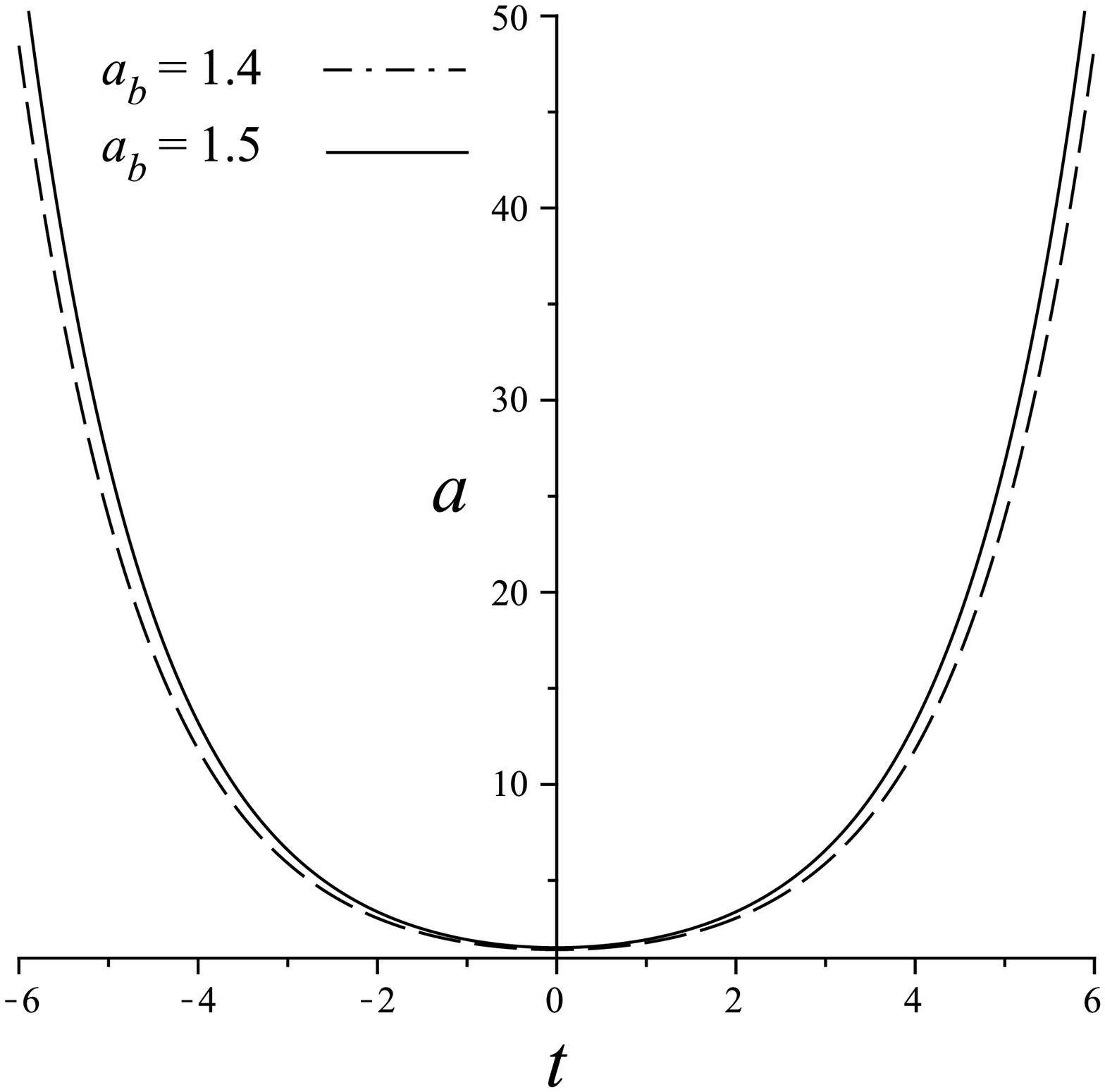}\hspace{0.1 cm}\includegraphics[scale=.3]{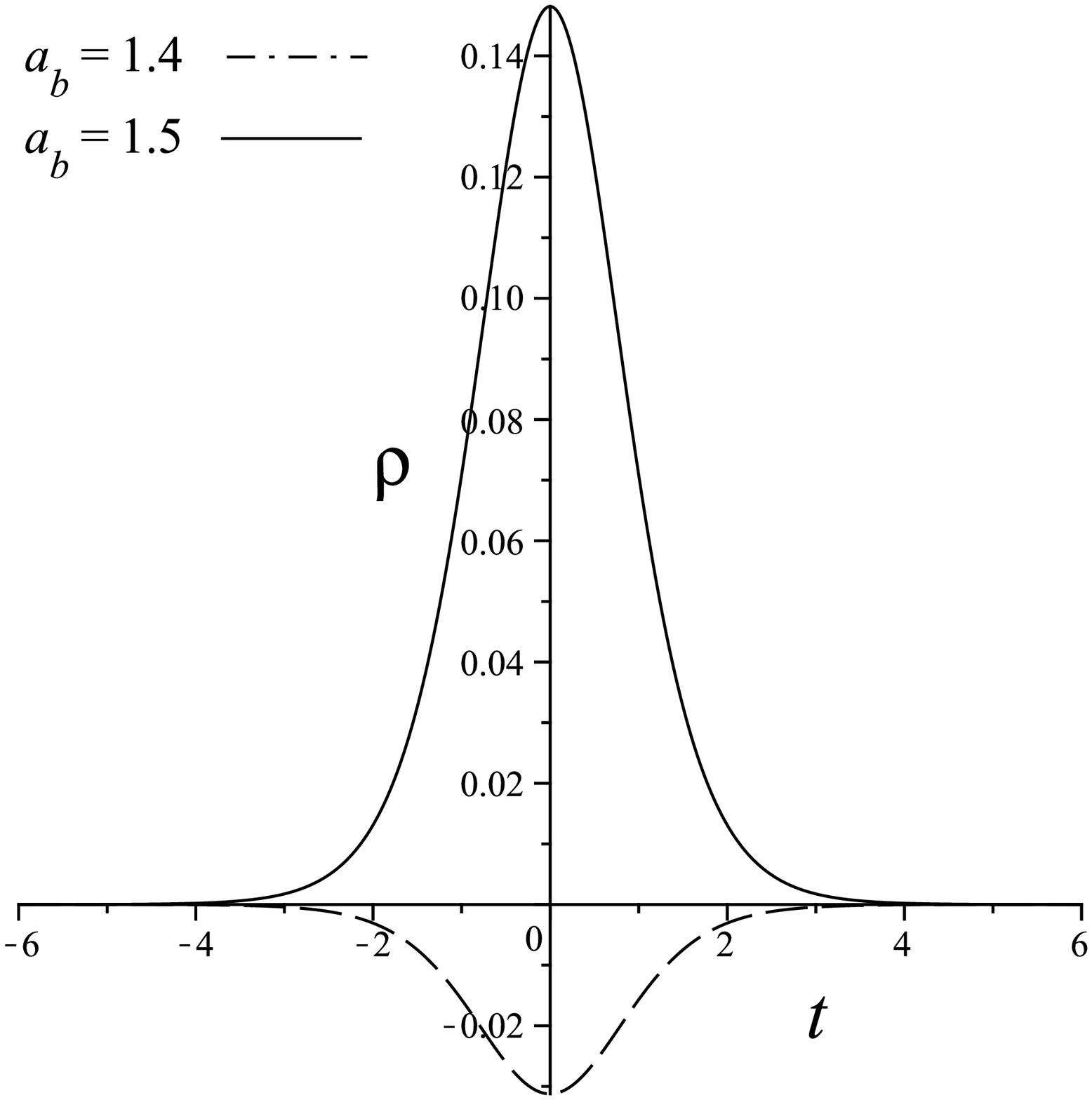}\hspace{0.1 cm}\\
Fig. 9:  Phase space of the dynamical system for the case  $k=1,\alpha=-2,\beta=0$ \\
\end{tabular*}\\

\subsubsection{$k=1,\alpha=0,\beta=positive$}
In this case, the system does not have any real critical point and does not oscillate. However.  minimal  condition for a bounce  may be satisfied. In this regard, given the positive $\rho_{b}=\frac{3(\zeta_{b}^{4}-\beta)}{\zeta_{b}^{6}}$ and $\frac{d\chi}{dt}|_{t_{b}}=-\frac{(\zeta_{b}^{4}+3\beta)}{2\zeta_{b}^{2}(\zeta_{b}^{4}-3\beta)}$   and  hence $\beta^{\frac{1}{4}}<\zeta_{b}<(3\beta)^{\frac{1}{4}}$, which may be obtained by setting $\alpha=0$ in equation (\label{in2}), there is a single bounce with positive energy density.

\subsubsection{$k=1,\alpha=0,\beta=negative$}

In this case, the system has two real critical points as $(\chi_{c}=0,\zeta_{c}=(-3\beta)^{\frac{1}{4}})$ and $(\chi_{c}=0,\zeta_{c}=-(-3\beta)^{\frac{1}{4}})$. the   corresponding eigenvalues are $\lambda_{1}=(\frac{i}{\zeta_{c}},\frac{-i}{\zeta_{c}})$ and $\lambda_{2}=(\frac{i}{\zeta_{c}},\frac{-i}{\zeta_{c}})$  which are the same.
The eigenvalues are purely imaginary and conjugated, as the critical points are centers (nonhyperbolic critical points) with closed curves turns around them. Consequently, the phase plane diagram in Fig. (10) simplifies the understanding of system as describing the
dynamics of cosmological parameters (scale factor and Hubble parameter), to provide a more meaningful insight on
the setting of initial conditions.

\begin{tabular*}{2.5 cm}{cc}
\includegraphics[scale=.42]{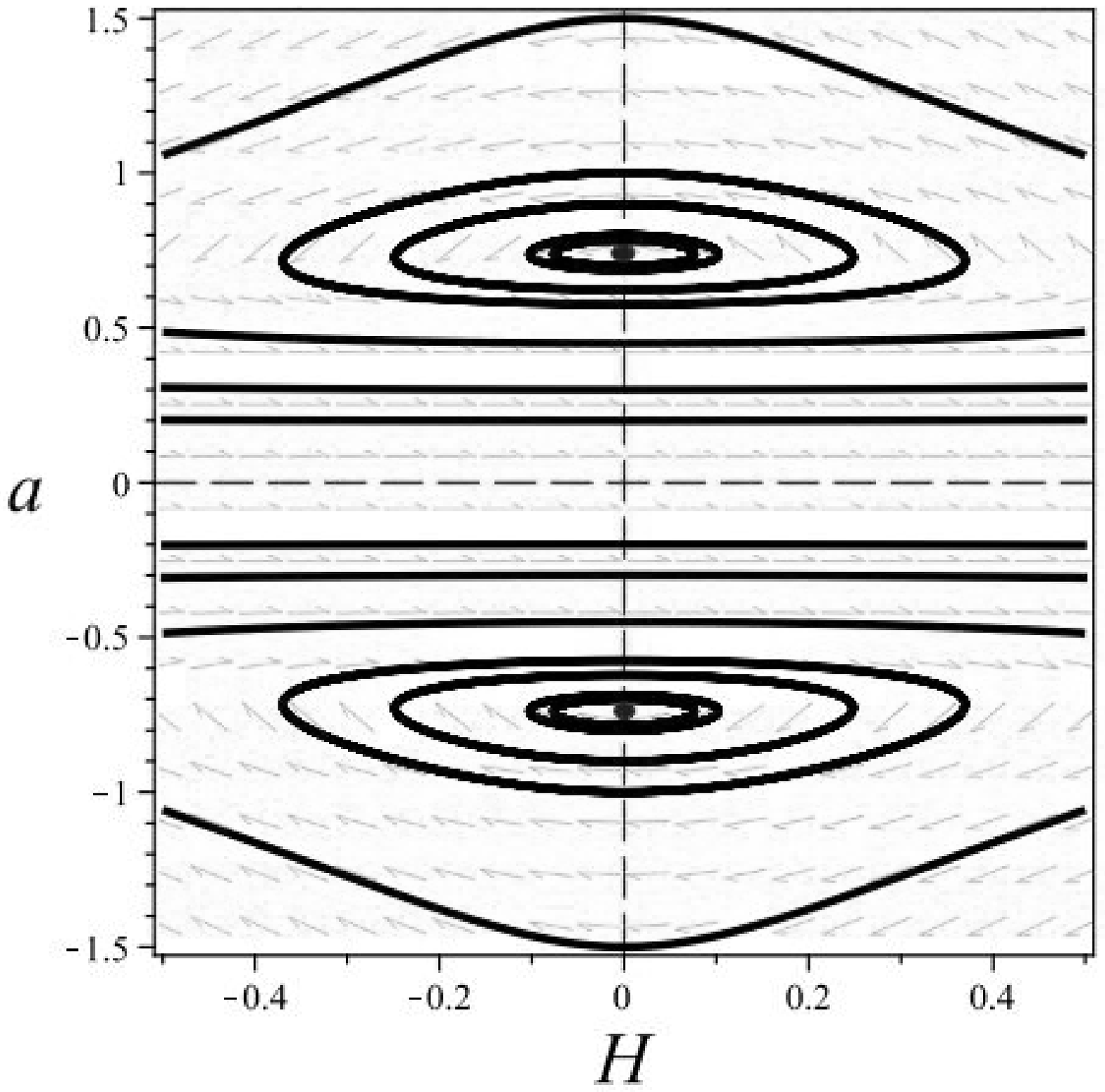}\hspace{0.1 cm}\\
Fig. 10:  The two dimensional  phase plane for  parameters $(a,H)$.We have set  $\alpha=0$, $\beta=-0.1$ . \\
\end{tabular*}\\

 \subsection{Case of hyperbolic ($k=-1$) universe}

By inclusion of a negative curvature index   $k=-1$ in modified Friedmann equations ((2) and (3)), the  number of critical points are as follows
\begin{equation}
P_{1}=\chi_{c}=0,\zeta_{c}=\frac{\sqrt{2}}{2}\sqrt{-\alpha+\sqrt{\alpha^{2}-12\beta}},\ \ \
P_{2}=\chi_{c}=0,\zeta_{c}=-\frac{\sqrt{2}}{2}\sqrt{-\alpha+\sqrt{\alpha^{2}-12\beta}},\\ \nonumber
\end{equation}
\begin{equation}
P_{3}=\chi_{c}=0,\zeta_{c}=\frac{\sqrt{2}}{2}\sqrt{-\alpha-\sqrt{\alpha^{2}-12\beta}},\ \ \
P_{4}=\chi_{c}=0,\zeta_{c}=-\frac{\sqrt{2}}{2}\sqrt{-\alpha-\sqrt{\alpha^{2}-12\beta}},\\ \nonumber
\end{equation}

Also,

\begin{equation}\label{rx}
\rho_{b}=-\frac{3(\zeta_{b}^{4}-\alpha\zeta_{b}^{2}-\beta)}{\zeta_{b}^{6}},\ \ \ \ \
h_{b}=\frac{(\zeta_{b}^{4}+\alpha\zeta_{b}^{2}+3\beta)}{2\zeta_{b}^{2}(\zeta_{b}^{4}-2\alpha\zeta_{b}^{2}-3\beta)}
\end{equation}

Given the minimal condition that  a bounce needs to occur from the local point of view while the energy density is positive $(h_{b}>0 ,\rho_{b}>0)$ [29], the various possibilities to satisfy this condition  can be classified as

                           $\left\{
\begin{array}{ll}
 \rho_{b}>0\Longrightarrow (\zeta_{b}^{4}-\alpha\zeta_{b}^{2}-\beta)<0 \\
 h_{b}>0\Longrightarrow\left\{
\begin{array}{ll}
 (\zeta_{b}^{4}+\alpha\zeta_{b}^{2}+3\beta)>0,(\zeta_{b}^{4}-2\alpha\zeta_{b}^{2}-3\beta)>0 \\
 \\
 (\zeta_{b}^{4}+\alpha\zeta_{b}^{2}+3\beta)<0,(\zeta_{b}^{4}-2\alpha\zeta_{b}^{2}-3\beta)<0\\
\end{array}
\right.
\end{array}
\right.
$\\
     \\

The above classification can be rearranged as two  sets of following conditions,
\\

                               CI:$\left\{
\begin{array}{ll}
 i:-(\zeta_{b}^{4}-\alpha\zeta_{b}^{2}-\beta)>0 \\
 ii:(\zeta_{b}^{4}+\alpha\zeta_{b}^{2}+3\beta)>0\\
 iii:(\zeta_{b}^{4}-2\alpha\zeta_{b}^{2}-3\beta)>0
\end{array}
\right.
$
CII:$\left\{
\begin{array}{ll}
 i:(\zeta_{b}^{4}-\alpha\zeta_{b}^{2}-\beta)<0 \\
 ii:(\zeta_{b}^{4}+\alpha\zeta_{b}^{2}+3\beta)<0\\
 iii:(\zeta_{b}^{4}-2\alpha\zeta_{b}^{2}-3\beta)<0
\end{array}
\right.
$\\
     \\

All of the (i), (ii) and (iii) conditions of CI class or all of them for CII class should be satisfied to have a bounce. The combining  of (i) and (ii) conditions of CI, gives the  $(2\alpha\zeta_{b}^{2}+4\beta)>0$ condition which is in contradiction  with the result of the combined provision of the (i) and (iii) conditions, i.e. $(2\alpha\zeta_{b}^{2}+4\beta)<0$.
However, a similar calculation on  the (i), (ii) and (iii) conditions of CII shows that it would be satisfied only if $\alpha>0$ and $\beta<0$.
One can conclude, therefore, that  there is the possibility of having the oscillating solution only for positive $\alpha$ and negative  $\beta$  of a negative curvature $k=-1$ universe.

\subsubsection{$k=-1,\alpha=positive,\beta=negative$}

For positive $\alpha$ and negative $\beta$, the expression $\alpha^{2}-12\beta>0$  and hence $(-\alpha+\sqrt{\alpha^{2}-12\beta}) $ are positive while $(-\alpha-\sqrt{\alpha^{2}-12\beta}) $ is negative.
 Consequently, there are only two of the above mentioned critical points, $P_{1}$ and $P_{2}$. The eigenvalues of the system can be real or complex.
 The three dimensional plot of $\Omega$, shown in Fig. (11) as a function of $\alpha$ and $\beta$, illustrates the positive and negative values of  $\Omega$  in different regions.

 For example by setting $\alpha=0.7$ and $\beta=-0.1$ the eigenvalues of system would be $\lambda=\pm(6.6I,-6.6I)$ (Fig. (12)), whereas for $\alpha=0.3$ and $\beta=-0.1$ the eigenvalues are real as $\lambda=\pm(2.3,-2.3)$. Therefore, it is possible to have a bouncing-oscillating  solution by proper choosing of $\alpha$ and $\beta$.  \\
\begin{tabular*}{2.5 cm}{cc}
\includegraphics[scale=.42]{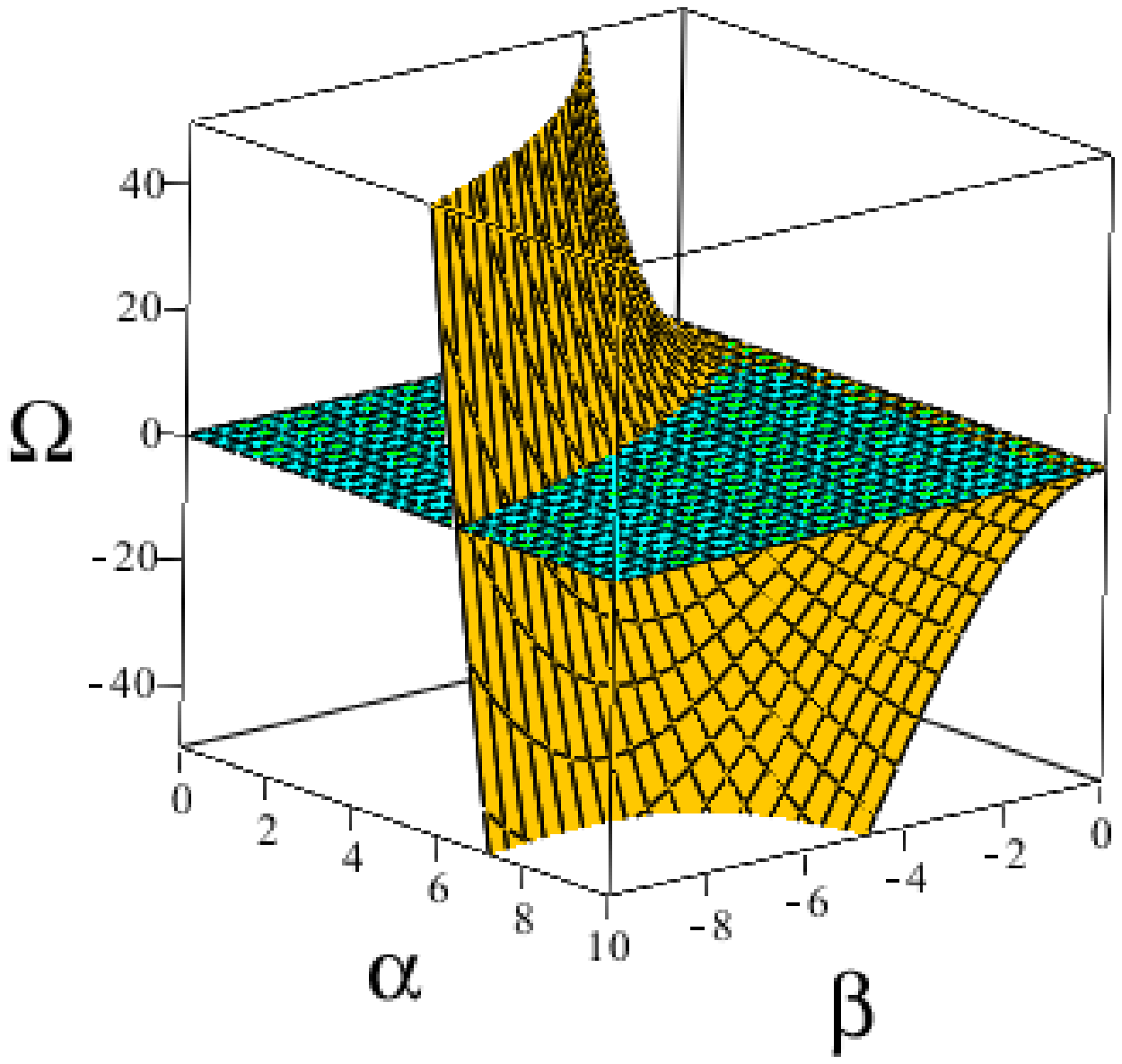}\hspace{0.1 cm}\\
Fig. 11:  The three dimensional   plot of $\Omega$ as a function of $\alpha$ and $\beta$ for $k=-1$.\\
\end{tabular*}\\
\begin{tabular*}{2.5 cm}{cc}
\includegraphics[scale=.6]{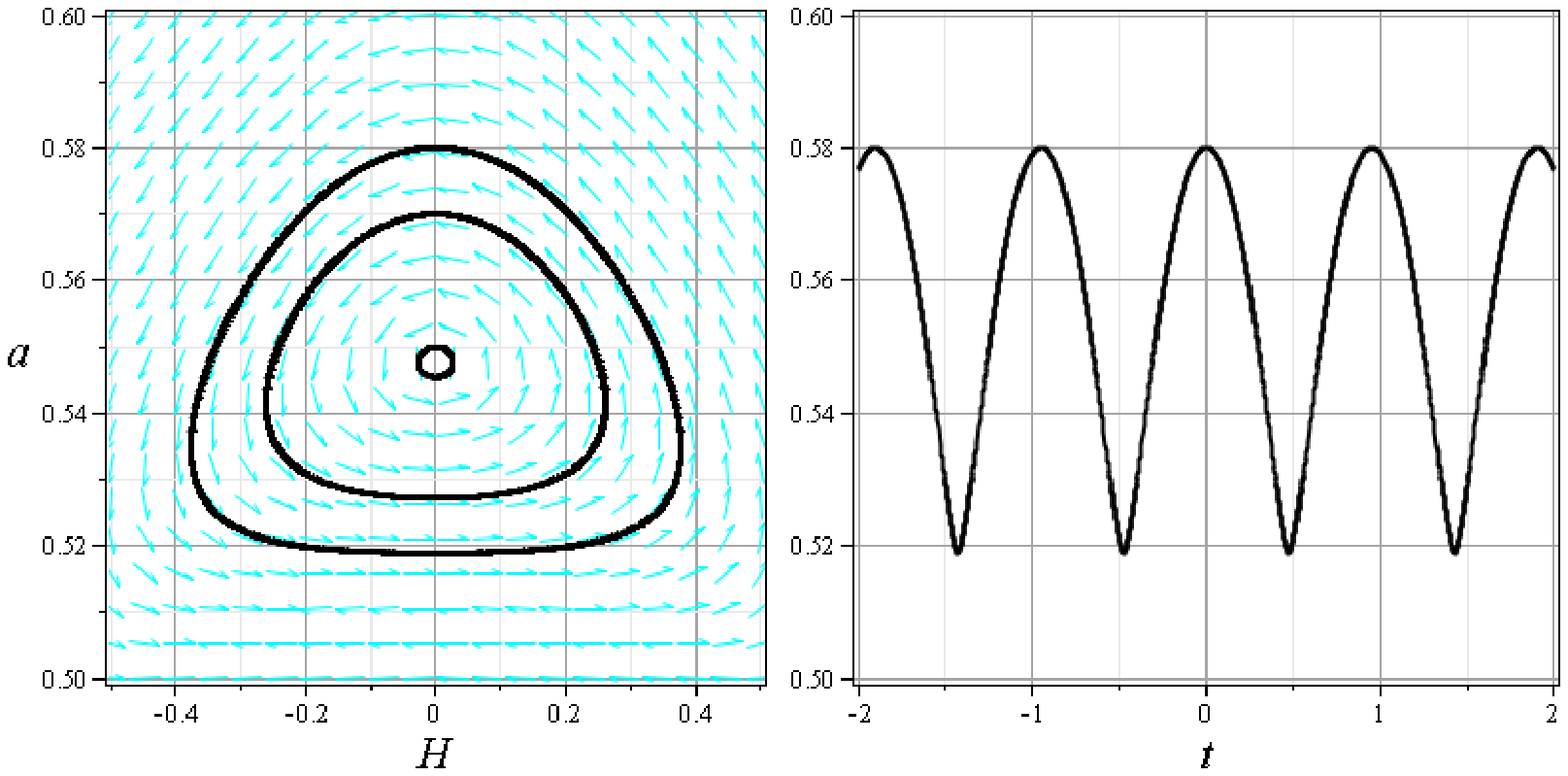}\hspace{0.1 cm}\\
Fig. 12:  The two dimensional  phase plane for  parameters $(a,H)$. We have set  $k=-1,\alpha=0.7$, $\beta=-0.1$. \\
\end{tabular*}\\

\subsubsection{$k=-1, Other cases$}

As it was mentioned above, other cases of a negative curvature    $k=-1$ universe have no oscillating solutions. The formal analysis appropriately can be applied the same as expressions of a positive curvature  universe. Let us consider the case of $\alpha=positive$ and $\beta=positive$, for example, where both the values of $(-\alpha-\sqrt{\alpha^{2}-12\beta})$ and $(-\alpha+\sqrt{\alpha^{2}-12\beta}) $ expressions are negative. Thus the system has no real critical point, indicating that the system has no oscillating solution. However, similar to some of the cases  described in the previous subsections, it is possible to have a single bounce if both the energy density and the derivative of hubble parameter at bounce  be positive. By tacking into account  these parameters at bounce, we have
\begin{equation}\label{rx}
\rho_{b}=-\frac{3(\zeta_{b}^{4}-\alpha\zeta_{b}^{2}-\beta)}{\zeta_{b}^{6}},\ \ \ \ \
\frac{d\chi}{dt}|_{t_{b}}=\frac{(\zeta_{b}^{4}+\alpha\zeta_{b}^{2}+3\beta)}{2\zeta_{b}^{2}(\zeta_{b}^{4}-2\alpha\zeta_{b}^{2}-3\beta)}
\end{equation}
Where positive energy density implies the negative $(\zeta_{b}^{4}-\alpha\zeta_{b}^{2}-\beta)$, and automatically the negative $\zeta_{b}^{4}-2\alpha\zeta_{b}^{2}-3\beta$. Hence the positive $\frac{d\chi}{dt}|_{t_{b}}$ requires the negative $(\zeta_{b}^{4}+\alpha\zeta_{b}^{2}+3\beta)$
which it is impossible for positive values of $\alpha$ and $\beta$, so  that there is no  bouncing solution to this case. One can see that the cases of $\alpha=positive,\beta=0$ and $\alpha=0,\beta=positive$ are the special forms of $\alpha=positive$, $\beta=positive$  with no bouncing solution.

The system has  two critical points, $P_{1}$ and $P_{2}$ for $\alpha=negative, \beta=negative$.   The energy density at bounce would be
 \begin{equation}\label{rx}
\rho_{b}=-\frac{3(\zeta_{b}^{4}-\alpha\zeta_{b}^{2}-\beta)}{\zeta_{b}^{6}}
\end{equation}
where for negative values of $\alpha$ and $\beta$ is negative. This means that without the aforesaid conditions a single bounce  never can occur. From equation \ref{omeg}, one can also understand this by noting the fact that the parameter $\Omega$ is positive and  says that the eigenvalues are real and no oscillating solution can exist. In addition, there are two of the above critical points, $P_{1}(\chi_{c}=0,\zeta_{c}=\sqrt{\alpha})$ and $P_{2}(\chi_{c}=0,\zeta_{c}=-\sqrt{\alpha})$ for the $\alpha=negative,\beta=positive$ case and also two critical points, $P_{1}(\chi_{c}=0,\zeta_{c}=\sqrt{\alpha})$ and $P_{2}(\chi_{c}=0,\zeta_{c}=-\sqrt{\alpha})$ for $\alpha=negative,\beta=0$. Here $\Omega=\alpha^{4}$ is positive  and indicates that the eigenvalues are real and no oscillating solution can exist.

\begin{table}[ht]
\caption{Number of Critical Points for $\alpha^{2}>12\beta$}   % title of Table
\centering % used for centering table
\begin{tabular}{|c|c|c|c|c|c|} % centered columns (5 columns)
\hline\hline %inserts double horizontal lines
Model  &  $\alpha$  & $\beta$  & Number of Critical Points& bounce& oscillation \\% inserts table
%heading
\hline % inserts single horizontal line
k=1 & positive & positive & 4 & Exist & yes \\ % inserting body of the table
\hline % inserts single horizontal line
k=1 &  positive & negative & 2  & Exist & yes\\
\hline % inserts single horizontal line
k=1 &  negative & positive & 0 & Exist & no \\
\hline % inserts single horizontal line
k=1 &  negative & negative & 2  & Exist & yes\\
\hline % inserts single horizontal line
k=1 &  0 & 0 & 1 & Don't Exist & no\\
\hline % inserts single horizontal line
k=1 &  0 & negative & 2  & Exist & yes\\
\hline % inserts single horizontal line
k=1 &  0 & positive & 0 & Exist & no \\
\hline % inserts single horizontal line
k=1 &  positive & 0 & 2  & Exist & yes\\
\hline % inserts single horizontal line
k=1 &  negative & 0 & 0 & Exist & no \\
\hline % inserts single horizontal line
k=-1 &  positive & negative & 4 & Exist & yes \\
\hline % inserts single horizontal line
k=-1 &  others & others& 4  & Don't Exist & no\\
\hline % inserts single horizontal line
\hline % inserts single horizontal line
\end{tabular}
\label{table:1} % is used to refer this table in the text
\end{table}\

\section{Conclusion}

The beginning of the universe in the context of general relativity without encountering singularities is not possible, as shown using the singularity theorem if a certain condition is satisfied. Oscillating universes have been explored to solve some problems of the standard cosmological model to avoid the big bang singularity and replace it with a cyclic evolution. Because of that, the outcome of bouncing models is very dependent on the choice of perturbation mechanisms in the background displaying a bounce; the aim of some specific models can be useful for extracting characteristics of a general expected behavior.

Since it is expected that by inclusion of quantum corrections, therefore, the singularity problem be replaced  by a quantum bounce; and also with cognizance of the thermodynamical area law that provides an important viability test for any theory of quantum gravity. The aim of this paper was twofold: exploring the bouncing solutions for modified Friedmann equations in order to find out the quantum entropy-corrected effects on the metric of FRW universe, and  possible signatures of its unknown parameters ($\alpha$ and $\beta$)  in satisfying the expected oscillatory evolutions (for matter dominated universes). At this point, from the consensus viewpoint, the value of the logarithmic pre-factor ($\alpha$) is a point of notable controversy which is constrained to be a negative (from the considerations of LQG) or positive (employed by statistical arguments) value and/or the ``best guess" might simply be zero.

By virtue of this matter, there is only the possibility of bouncing/oscillating solutions for curved universes ($k\neq 0$). In this right, there is the   possibility of oscillating solutions and a single bounce (a bouncing evolution
without regular repetition) under some appropriate conditions  for positive curvature $k=1$  universe with positive $\alpha$ and negative $\alpha$,  respectively (with any given value of $\beta$); whereas a non-singular solution does not exist for the  case of  $k=1$ universe with   $\alpha=0$ (and $\beta=0$). Furthermore, the possibility of having the oscillating solutions there is only for positive $\alpha$ (and negative  $\beta$)  of a negative curvature $k=-1$ universe. As a result of what was said, accordingly, since the cyclic evolutions should be appeared in discussing the quantum corrections to Friedmann equations; we can emphasize the necessity of non-zero $\alpha$ and by the way recommend  the positive values of  $\alpha$ for the possibility of having the oscillating solutions and negative values of $\alpha$  for the presence of a quantum bounce. In addition to this result, a detailed description of calculations and how to find the stability analysis of systems was performed in the paper, where illustrates  the direction of handling a modified gravity theory along with phase plane analysis in discussing the bouncing evolution of cyclic scenario.

\end{document}